\documentstyle[12pt]{article}
\begin{document}
\def\theequation{\arabic{section}.\arabic{equation}}
\def\thesection{\arabic{section}}
\renewcommand{\thefootnote}{\fnsymbol{footnote}}
\setcounter{footnote}{2}

\section{Introduction}

\setcounter{equation}{0}

In a recent letter we have presented the bare
 finite temperature effective
potential for regularized $SU(2)$ Yang-Mills theory taking
the Haar measure into account\cite{Sai95}. Here we present the
renormalized effective potential at the order $g^2$
and show that it develops a non-perturbative minimum
for sufficiently strong coupling, i.e. below a critical
temperature. We discuss that this minimum corresponds a phase
which can be a candidate for the confining phase of the
continuum theory.

   As to confinement in pure
Yang-Mills theories general arguments were given that the
 deconfining
phase transition occurs with increasing temperature if the theory
 is confining at zero temperature \cite{Pol78,Sus79}.
 It was also shown that
lattice gauge theory does not confine static quarks if the
 Haar measure
is replaced by the Euclidean one \cite{Pat81}.
  Non-trivial phases for
$SU(2)$ continuum theory has  been predicted in Ref. \cite{Boh94}.

The 2--loop contribution of the order $g^2$
 to the effective potential of $SU(N)$ gauge theory is
well-known in the perturbative
regime  \cite{Kap79,Gro81}.  Now
we determine the effective potential for $SU(2)$  up to
this order by using the non-perturbative mean-field approach
and treating the fluctuations around the mean field perturbatively.
This mixture of the non-perturbative and perturbative approaches
provides some insight in the mechanism producing
 the non-perturbative
phase below the critical temperature.
We assume that
 the component
 $A^{\mu =0 \; a=3}$ ($\mu$ Lorentz index, $a$ $SU(2)$ colour
index) of the gluon vector potential
has the non-vanishing  expectation value,
 $v \equiv \langle a^{-1} \beta g A^{03}
\rangle$
 ($a$ the lattice spacing for the regularized path integral,
$\beta$ the inverse temperature).
 Doing so we cannot assume any more that the field $A^{03}$
 is small
and,   consequently, we must not replace the Haar measure,
\begin{eqnarray}
      \prod_{ {\vec x} } d^3 \alpha^{a}_{\vec x}
      \frac{ \sin^2  \left( \frac{1}{2} (  \alpha^{a}_{\vec x}
              \alpha^{a}_{\vec x} \; )^{1/2}    \right)
          }{
          \alpha^{a}_{\vec x}  \alpha^{a}_{\vec x}
                }
\end{eqnarray}
(with $\alpha_{\vec x}^a = g \beta a^{-1} A^{0a}_{\vec x}$,
\cite{Polo90,Joh91})
by the Euclidean one in the path integral. This is in  contrast to
that
one generally does in
the perturbative approach.

Lattice results have shown that confinement and global center
symmetry are strongly related \cite{Polo90}. Since global center
 symmetry of the path integral is guaranteed by the usage of the
 Haar measure, the question arises
 whether  does the Haar measure influence the effective potential.
There are general arguments that it does not influence the physics
in the  perturbative regime \cite{Zin89,Rei88}.
It was shown for $SU(2)$ Yang-Mills theory that
the tree level contribution of the Haar measure
is cancelled by a piece of the 1-loop contribution of
 the longitudinal
gluons \cite{Wei81}.  Thus the Haar measure does not play
 any role in
the effective potential at the order $g^0$ even if
a finite background field is assumed.
In \cite{Sai95} we have shown that the Haar measure reveals itself
in the bare effective potential
of the pure $SU(2)$ Yang-Mills theory
at the order $g^2$. The authors of
 \cite{Boh94} found a cancellation of the Haar measure induced
 terms
using a different approximation. Later we discuss the
 possible reason
of this disagreement. Nevertheless,
our results for the bare theory are in qualitative agreement
with the
findings for
lattice gauge theory \cite{Pol82} as discussed in \cite{Sai95}.
Here we show that the basic features of the renormalized effective
 potential are not influenced by the Haar measure. It can reveal
itself only in the fine details of the effective potential (in
the approximation used).

Our strategy of treating the IR and UV divergences is as follows.
The bare theory is formulated in terms of three dimensional
 parameters:
the inverse temperature $\beta$, the UV momentum cut-off $\Lambda$,
and the IR momentum cut-off $\mu$. Thus UV divergences can occur
as powers and logarithms of $\beta \Lambda$ and $\Lambda /\mu$.
UV divergences of the first type are those controlled by power
counting
and are removed by subtracting loops in the
 limit $\beta \to \infty$
\cite{Kapbook}. Of the second type UV divergences
 accompanying the IR
divergences  need some particular treatment. Their counterterms are
introduced by choosing the IR momentum cut-off as a power series
of $1/\Lambda$ and the logarithmically divergent
 terms are neglected.
This removes both IR and UV divergences.
 The IR singularities  are now
 removed in a way getting independent of the
UV cut-off $\Lambda$ when approaching the continuum limit
 (the rather
correct approach instead of our earlier
one \cite{Sai95}), since the IR problem in the real world
 is solved via dynamical
generation of electric and magnetic masses, defining a physical
mass scale independent of the UV cut-off.

In \cite{Wei81} the effective potential was determined at the order
$g^0$ allowing for the non-vanishing vacuum expectation value $C/g
\equiv ( a/\beta )(v/g)$.
In the present paper we determine
 the terms of the order $g^2$ of the effective potential
 by using the
general techniques of finite temperature field theory
 \cite{Kapbook}.
Following \cite{Wei81} we
use a time independent, diagonal gauge,
and periodic boundary conditions for the spatial components of the
vector potential
 in the `time' direction. On the other hand we take all the terms
of the order $g^2$ with in the effective potential.
In this gauge the Haar measure induced potential in the tree level
action takes the form:
\begin{eqnarray}
     V_{H \; tree} &=& - \frac{1}{a^3} \int d^3 x
      \ln \left(  1 - \cos \alpha^3 ( {\vec x} )   \right) .
\end{eqnarray}

 Our discussion starts with establishing
the 2-loop corrections in terms of the gluon propagators. Then we
determine their explicit expressions revealing their
 dependence on the
IR momentum cut-off $\mu$ and the UV momentum cut-off $\Lambda$,
and
perform the first step of the renormalization procedure by
 subtracting
the corresponding loops at zero temperature.
 As to the next we complete
the renormalization procedure by
 removing
the additional UV divergences together with the IR divergences.
Finally, we discuss the phase structure of the continuum $SU(2)$
Yang-Mills theory on the base of the renormalized
 effective potential obtained.

 Throughout this paper we use the notations of
 Ref. \cite{Wei81} and use the same cut-off regularization. The UV
cut-off $\Lambda$ is interpreted in terms of the lattice
 spacing $a$
used for the definition of the path integral via
$a^{-3} = (2\pi )^{-3} \int_0^\Lambda d^3 k $ with $\Lambda =
(6\pi^2 )^{1/3} a^{-1}  $ and $d^3 k$ the volume element in
3-momentum space. The same spacing $a$ is assumed in the `time'
direction.

\section{The bare effective potential at 2-loop order}

\setcounter{equation}{0}

Let us allow  for the non-vanishing  expectation value
$(a/\beta )(v/g)$ of the component $A^{03}$ of the gluon vector
potential and shift the field variable by it.
Inserting $A^{03} ({\vec x} ) =(a/\beta ) (v /g)
+ \delta \phi ( {\vec x} )$ in the Euclidean action
\begin{eqnarray}
S_{YM} &= &   - \frac{1}{4a^4 \hbar}
    \int_0^{\beta } dx^0 \int d^3 x  F^{\mu \nu \; a} F^{\mu \nu \;
       a}
\end{eqnarray}
of the Yang-Mills system ($F^{\mu \nu \: a}$
 the field strength tensor),
the tree level action takes the following form:
 \begin{eqnarray}
     S_{tree} &=& S_0 + S_1 + S_2 .
\end{eqnarray}
 The term
\begin{eqnarray}
\label{S0}
   S_0 &=&
   \frac{1}{2a^4 \hbar} \int_0^\beta dx^0  \int d^3 x \; a^2
   \left\lbrack
     ( \nabla^i \delta \phi )^2
       + \left(  \partial^0 A^{ia} + (v/\beta )
         \epsilon^{a3b} A^{ib} \right)^2
            \right.
                   \nonumber\\
        &  &  \left. \qquad \qquad
     + \frac{1}{2} ( \partial^i A^{ja} - \partial^j A^{ia} )^2
   \right\rbrack
          \nonumber\\
    &  &  -
    \frac{1}{a^3} \int d^3 x \ln ( 1 - \cos v )
   + \frac{1}{2 a^3 } \frac{ g^2 a^{-2} \beta^2 }{ 1 - \cos v }
      \int d^3 x (\delta \phi )^2
\end{eqnarray}
 contains the action of the free gluon field of
the order $(g \sqrt{\hbar} )^0$  and the
Haar measure induced self-interaction of timelike gluons,
a term  of order $g^2 \hbar$ (the fields are of order $\hbar^{1/2}$).
The other Haar measure induced vertices are neglected.
$S_1$ and $S_2$ contain the usual 3-gluon and 4-gluon interaction
vertices:
\begin{eqnarray}
\label{S1}
      S_1 &=& \frac{g}{a^4 \hbar} \int d^4 x   \; a
  \left\{  \epsilon^{a3d} \delta \phi \;  A^{id}
    \left( \partial^0 A^{ia} + (v/\beta ) \epsilon^{a3c} A^{ic}
\right)
    +  \epsilon^{abc} A^{ib} A^{jc}
      \partial^i A^{ja}       \right\}  ,
             \nonumber\\
\end{eqnarray}
\begin{eqnarray}
\label{S2}
     S_2 &=& \frac{g^2}{2a^4 \hbar}
      \int d^4 x \left\{  \epsilon^{3ca} \epsilon^{3da}
      A^{ic} A^{id} (\delta \phi )^2
  +    \frac{1}{2} \epsilon^{abc} \epsilon^{ade} A^{ib} A^{jc} A^{id}
              A^{je}  \right\}
\end{eqnarray}
($\epsilon^{abc}$ the completely antisymmetric tensor of rank 3).
The self-interaction terms $S_1$ and $S_2$ are of the order
$g\hbar^{1/2}$ and $g^2 \hbar$, respectively.

The main goal of this section is to determine the effective potential
as function of the constant background field $v$
 including all the terms being of the same order $g^2$
  as the one-loop contribution from the Haar measure. Then the
interaction $S_1$ and $S_2$ must be taken into account at 2-loop level.
Our procedure is the following:
\begin{itemize}
 \item  determine the partition function by
treating the  interaction terms $S_1$ and $S_2$  perturbatively;
 \item determine the effective potential from the 1PI diagrams.
\end{itemize}

In order to calculate the perturbative corrections
we introduce
the partition function in the presence of the external sources
$j^{ia}$ and $q$:
\begin{eqnarray}
        {\cal Z} \lbrack j^{ia} , q \rbrack &=&
     \exp \left\{  - {\hat S}_1  - {\hat S}_2  \right\}
       {\cal Z}_0 \lbrack j^{ia} ,q \rbrack
\end{eqnarray}
with the generating functional of the free fields
\begin{eqnarray}
      {\cal Z}_0 \lbrack j , q \rbrack
   &=&  \int {\cal D} \delta \phi {\cal D} A^{ia}
   e^{ - S_0 - (jA)  - (q \delta \phi )  }
    =  {\cal Z}_0 \lbrack 0,0 \rbrack
       z \lbrack j, q \rbrack ,
\end{eqnarray}
where
\begin{eqnarray}
        z \lbrack j, q \rbrack &=&
       \exp \left\{  \frac{1}{2} ( j D j )
                 + \frac{1}{2}  ( q {\cal D} q )
                  \right\}
\end{eqnarray}
and
\begin{eqnarray}
   {\cal Z}_0 \lbrack 0,0 \rbrack &=&
    \left(  {\mbox{Det}} {\cal M}  \cdot
            {\mbox{Det}} ( {\cal N} + {\cal N}_H  )
    \right)^{-1/2}
     \exp \left\{ a^{-3} \int d^3 x \ln ( 1 - \cos v )    \right\} .
                \nonumber\\
\end{eqnarray}
Here $D = {\cal M}^{-1}$ and
 ${\cal D} = ( {\cal N} + {\cal N}_H )^{-1}$
are the free propagators of the spatial and timelike gluons,
respectively,
defined through
\begin{eqnarray}
     S_0 &=& - a^{-3} \int d^3 x \ln ( 1 - \cos v )
   +   \frac{1}{2} ( A {\cal M} A )
          +     \frac{1}{2} ( \delta \phi \; ( {\cal N} + {\cal N}_H )
                      \;       \delta \phi ) .
              \nonumber\\
\end{eqnarray}
The matrices $D$ and ${\cal M} $ (${\cal D}$ and ${\cal N}$, ${\cal
N}_H$) possess Lorentz, colour and spacetime coordinate (spatial
coordinate) indices, $(\ldots )$ is a shorthand for contraction.
The operator ${\hat S}_{int} = {\hat S}_1 + {\hat S}_2$ is defined by
replacing the fields $A^{ia}$ and $\delta \phi$ by functional
derivatives
\begin{eqnarray}
        A^{ia}  \to - \frac{ \delta}{ \delta j^{ia} }  ,
        \qquad
        \delta \phi \to - \frac{ \delta }{ \delta q }
\end{eqnarray}
with respect to the corresponding external sources in the
expressions for $S_1$ and $S_2$.

We expand the exponential operator in powers of $(g \hbar^{1/2} )$:
\begin{eqnarray}
     e^{ -{\hat S}_{int}  } &=&
      1 - {\hat S}_1  + \left\lbrack
        - {\hat S}_2
     + \frac{1}{2} ( {\hat S}_1 )^2 \right\rbrack  +
    {\cal O} \left( ( g \sqrt{\hbar} ) \right)^3  .
\end{eqnarray}
After some algebra we obtain ($\hbar =1$):
\begin{eqnarray}
       {\hat S}_1 z \lbrack j,q \rbrack &=&  -
      g \int dx \; a \left\{ \epsilon^{3da} ( {\cal D}q )
     \left\lbrack (Dj )^{id}
     \left( \partial^0 (Dj)^{ia} + (v/\beta ) \epsilon^{3ca} (Dj)^{ic}
\right)
         \right.   \right.
                \nonumber\\
      &  &   \qquad \qquad \left.
    + \partial_x^0 D^{iiad}_{xx'} + (v/\beta ) \epsilon^{3ca}
D^{iidc}_{xx}
      \right\rbrack
                  \nonumber\\
        &  & \qquad  + \epsilon^{abc}
      \left\lbrack (Dj)^{ib} ( Dj)^{jc} \partial^i_x (Dj)^{ja}
       +   D^{ijbc}_{xx} \partial^i (Dj)^{ja}
           \right.      \nonumber\\
       & & \left.  \left.
     \qquad \qquad  + (Dj)^{ib} \partial_x^i D^{jjac}_{xx'}
       + (Dj)^{jc}  \partial^i_x D^{jiab}_{xx'}
       \right\rbrack    \right\}
         z \lbrack j, q \rbrack ,
\end{eqnarray}
\begin{eqnarray}
\label{S1z}
  \left.   {\hat S}_1 z \lbrack j,q \rbrack
   \right|_0  =0 ,
\end{eqnarray}
and
\begin{eqnarray}
     \left.  {\hat S}_{2} z \lbrack j, q \rbrack  \right|_{0} &=&
     \frac{1}{2} g^2 \int dx \left\{   \epsilon^{3ca}
\epsilon^{3da}
      {\cal D}_{xx}  D_{xx}^{iicd}
                  \right.
             \nonumber\\
        &  &  \left.
        +  \epsilon^{abc} \epsilon^{ade}
       \left( D^{ijbc}_{xx} D^{ijde}_{xx}
            + D^{ijbe}_{xx} D^{jicd}_{xx}
            + D^{iibd}_{xx} D^{jjec}_{xx}  \right)
             \right\}  .
\end{eqnarray}
The 1PI part of ${\hat S}_1 ( {\hat S}_1 z )   |_0$ can
easily be identified by
noticing that (i) the terms containing first or third  functional
derivatives of $z \lbrack j^{ia} , q \rbrack$
 with respect of the external sources  vanish, (ii)
the
non-vanishing terms with  second functional derivatives of
  $z \lbrack j^{ia} , q \rbrack$  are
1-particle reducible. Thus the terms not containing the derivatives of
$z$ are the only ones contributing to the 1PI part:
\begin{eqnarray}
   \left. \frac{1}{2} {\hat S}_1 ( {\hat S}_1 z ) \right|_{0 \;
     \mbox{1PI} }   &=&
     '{\cal D} D D '  \quad   + \quad   'DDD ' ,
\end{eqnarray}
where
\begin{eqnarray}
      {\mbox{ '}}{\cal D} DD '  &=&     \epsilon^{3da} \epsilon^{3eg}
    \frac{g^2}{2a^8} \int d^4 y d^4 x   {\cal D}_{xy}  \cdot
             \nonumber\\
     &  &  \cdot a^2 \left\lbrack
      D^{kied}_{xy} \partial^0_x \partial^0_y D^{kiga}_{xy}
   +  \partial^0_y D_{xy}^{kiea} \cdot \partial^0_x D^{kigd}_{xy}
             \right.      \nonumber\\
    & &
     +  (v/\beta ) \epsilon^{3fg} \left(
     D^{kied}_{xy} \partial^0_y D^{kifa}_{xy}
   +  D^{kifd}_{xy} \partial^0_y D^{kiea}_{xy}  \right)
              \nonumber\\
     &  &
     + (v/\beta ) \epsilon^{3ca} \left(
      D^{kied}_{xy} \partial^0_x D^{kigc}_{xy}
     + D^{kiec}_{xy} \partial^0_x D^{kigd}_{xy}  \right)
               \nonumber\\
     &  &   \left.
           +  (v/\beta )^2  \epsilon^{3ca} \epsilon^{3fg}
       \left( D^{kied}_{xy} D^{kifc}_{xy}
         +  D^{kiec}_{xy}  D^{kifd}_{xy}  \right)
            \right\rbrack ,
\end{eqnarray}
\begin{eqnarray}
   ' DDD ' \quad  &=&
    \frac{g^2}{2a^8} \epsilon^{abc} \epsilon^{gef}
    \int d^4 y d^4 x   a^2 \left\lbrack
      D^{kieb}_{xy} D^{ljfc}_{xy} \partial^i_y \partial^k_x
                D^{ljga}_{xy}
              \right.     \nonumber\\
      &  &
   +   D^{kieb}_{xy} \partial^i_y D^{ljfa}_{xy} \cdot \partial^k_x
                D^{ljgc}_{xy}
   +   D^{kjec}_{xy} D^{lifb}_{xy} \partial^i_y \partial^k_x
                D^{ljga}_{xy}
             \nonumber\\
       &  &
   +   D^{kjec}_{xy} \partial^i_y D^{ljfa}_{xy} \cdot \partial^k_x
                D^{ligb}_{xy}
   +   D^{lifb}_{xy} \partial^i_y D^{kjea}_{xy}  \partial^k_x
                D^{ljgc}_{xy}
            \nonumber\\
         &  &   \left.
    +  D^{ljfc}_{xy} \partial^i_y D^{kjea}_{xy} \partial^k_x
                D^{ligb}_{xy}
               \right\rbrack.
\end{eqnarray}

\begin{figure}

\vspace{8cm}

\caption{2--loop contributions to the effective potential.}
\end{figure}

The effective potential is defined by the 1PI part of the partition
function.
The 1-loop contribution of the gluon self-interaction $S_{int}$
 to the effective potential
vanishes identically due to Eq. (\ref{S1z}).
 The 2-loop contribution $\Delta V_{eff}$ is given by
\begin{eqnarray}
   {\cal Z} \lbrack 0,0 \rbrack _{\mbox{1PI}}
      &=& {\cal Z}_0 \lbrack 0,0 \rbrack
      \exp \{ - \beta V \Delta V_{eff} \}
\end{eqnarray}
with
\begin{eqnarray}
\label{DF}
    \Delta V_{eff} &=& (\beta V)^{-1} \left. {\hat S}_2 z \lbrack j,q
\rbrack
\right|_0
        -  \frac{1}{2\beta V } \left.  \left(  {\hat S}_1 \right)^2 z
\lbrack
              j,q  \rbrack
                  \right|_{0 \; \mbox{1PI} }
                \nonumber\\
        &  \equiv &
      f_1 + f_2  + f_3 + f_4
\end{eqnarray}
($V \to \infty $ the 3-volume of the system).
The various terms on the r.h.s. of Eq. (\ref{DF}) correspond to the
diagrams in Fig. 1
and represent the bare 2--loop contributions to the effective
potential:
\begin{eqnarray}
\label{f1B}
    f_{1B} &=&  \frac{g^2}{2a^4}
    \epsilon^{abc} \epsilon^{ade} \left(
            D^{ijbc}_{xx} D^{ijde}_{xx}
        +    D^{ijbe}_{xx} D^{jicd}_{xx}
        +    D^{iibd}_{xx} D^{jjec}_{xx}       \right) ,
\end{eqnarray}
\begin{eqnarray}
\label{f2B}
    f_{2B} &=&
   \frac{1}{2 \beta Va^4} g^2 \int d^4 x \left( \delta^{33} \delta^{cd}
      -  \delta^{3c} \delta^{3d}  \right)
     {\cal D}_{xx}  D^{iicd}_{xx}
                \nonumber\\
      &=&  \frac{g^2}{2a^4} {\cal D}_{xx} \left(
      D^{iicc}_{xx} - D^{ii33}_{xx}  \right)  ,
\end{eqnarray}
\begin{eqnarray}
\label{f3B}
    f_{3B} &=&
      \frac{1}{2\beta Va^8} g^2 \epsilon^{abc} \epsilon^{gef}
         a^2 \int d^4 y d^4 x
     \left\lbrack  D^{kieb}_{xy} D^{ljfc}_{xy}
       \partial^i_x \partial^k_x D^{ljga}_{xy}
                   \right.    \nonumber\\
      & &
    + D^{kieb}_{xy} \partial^i_x D^{ljfa}_{xy} \cdot
        \partial^k_x D^{ljgc}_{xy}
    + D^{kjec}_{xy} D^{lifb}_{xy} \partial^i_x \partial^k_x
        D^{ljga}_{xy}
                \nonumber\\
       &  &
     + D^{kjec}_{xy} \partial^i_x D^{ljfa}_{xy} \cdot \partial^k_x
         D^{ligb}_{xy}
                    \nonumber\\
       &  &
       \left.
     + D^{lifb}_{xy} \partial_x^i D^{kjea}_{xy} \cdot
        \partial_x^k D^{ljgc}_{xy}
     + D^{ljfc}_{xy}  \partial^i_x D^{kjea}_{xy} \cdot
        \partial^k_x D^{ligb}_{xy}
              \right\rbrack   ,
\end{eqnarray}
\begin{eqnarray}
\label{f4B}
      f_{4B} &=&
     - \frac{g^2}{2 \beta Va^8} \epsilon^{3da} \epsilon^{3eg}
      \int d^4 y d^4 x {\cal D}_{xy} \cdot
                     \nonumber\\
     &  &  \cdot a^2 \left\lbrack
     - D^{kied}_{xy} (\partial^0_x )^2 D^{kiga}_{xy}
     + D^{kiea}_{xy} (\partial^0_x )^2 D^{kigd}_{xy}
                \right.
               \nonumber\\
      &  &  \qquad
     + (v/\beta ) \epsilon^{3fg} \left( - D^{kied}_{xy}
         \partial^0_x D^{kifa}_{xy}
        +  D^{kiea}_{xy} \partial^0_x D^{kifd}_{xy}  \right)
                \nonumber\\
       &   &  \qquad
      + (v/\beta ) \epsilon^{3fg} \left( D^{kide}_{xy} \partial^0_x
             D^{kiaf}_{xy}
         + D^{kidf}_{xy} \partial^0_x D^{kiae}_{xy}  \right)
                \nonumber\\
       & &  \qquad  \left.
      + (v/\beta )^2 \epsilon^{3ca} \epsilon^{3fg}
      \left( D^{kied}_{xy} D^{kifc}_{xy}  +
             D^{kiec}_{xy} D^{kifd}_{xy}   \right)
                 \right\rbrack .
\end{eqnarray}
Later we shall discuss how to obtain the renormalized contributions
$f_1$, $f_2$, $f_3$, and $f_4$ from the corresponding bare ones.

The 2-loop corrections (\ref{f1B})-(\ref{f4B})
are given in terms of the free gluon
propagators.
After changing to momentum representation we read them off from the
free action $S_0$. Then we obtain for the free propagator of  timelike
gluons
\begin{eqnarray}
\label{calDk}
     {\cal D} ( {\vec k} ) &=&
     \left\lbrack  a^2  ( {\vec k}^{\; 2} + M^2 )  \right\rbrack^{-1}
\end{eqnarray}
with $M^2 a^2 \equiv a^{-1} g^2 \beta ( 1 - \cos v )^{-1} $ and
for that of the spatial gluons
\begin{eqnarray}
\label{spapro}
   D^{ijab} (\omega_n , {\vec p} )
      &=& a^{-2}
 \left\lbrack
 (\omega_n^2 + {\vec p}^{\; 2} ) \delta^{ij} \delta^{ab}
    + (v/\beta )^2 \delta^{ij} ( \delta^{ab}
    - \delta^{a3} \delta^{b3} )
             \right.
                  \nonumber\\
     &  &  \qquad  \left.
     - p^i p^j \delta^{ab} - 2 (v/\beta ) {\rm i} \omega_n
\epsilon^{3ab}
           \delta^{ij}
          \right\rbrack^{-1}
\end{eqnarray}
with the bosonic Matsubara frequencies $\omega_n =2\pi n /\beta$
$(n=0, \pm 1, \pm 2 \ldots )$.
Separating the transverse and the longitudinal parts
the matrix on the r.h.s. of Eq. (\ref{spapro}) can be inverted:
\begin{eqnarray}
\label{DFou}
  D^{ijab} (\omega_n , {\vec p} ) &=&   a^{-2}
  \Delta^{ab} (\omega_n , 0 )
  \frac{p^i p^j}{ {\vec p}^{\; 2}  }
   + a^{-2} \Delta^{ab} (\omega_n , {\vec p} )
  \left( \delta^{ij}  - \frac{ p^i p^j }{ {\vec p}^{\; 2}  } \right) ,
           \nonumber\\
\end{eqnarray}
with
\begin{eqnarray}
\label{Delab}
  \Delta^{ab} (\omega_n , {\vec p} ) &=&
       \left(  \!
    \begin{array}{ccc}
    \frac{1}{2} \left( d_n^+ ({\vec p } ) + d_n^- ({\vec p} ) \right)
  &  - \frac{\rm i}{2} \left( d_n^+ ({\vec p} ) - d_n^- ({\vec p} )
              \right)            &   0      \\
  + \frac{\rm i}{2} \left( d_n^+ ({\vec p} ) - d_n^- ({\vec p} )
       \right)     &
    \frac{1}{2} \left( d_n^+ ({\vec p } ) + d_n^- ({\vec p} ) \right)
               &  0     \\
    0   &   0  &  d_n ( {\vec p} )
     \end{array}   \!   \right)
             \nonumber\\
\end{eqnarray}
and
 $d_n^\pm ( {\vec p} ) = \left\lbrack ( \omega_n^\pm )^2 + {\vec
p}^{\; 2}   \right\rbrack^{-1}  $, $d_n ( {\vec p} ) =
 \left( \omega_n^2
 + {\vec p}^{\; 2}  \right)^{-1} $,
 $\omega_n^\pm = \omega_n \pm (v/\beta ) $.
In coordinate representation, the free propagators are then given by
\begin{eqnarray}
\label{calDcor}
    {\cal D}_{ {\vec x}{\vec y} }
    &=& a^3 \int \frac{ d^3 k}{ (2\pi )^3 }
    \frac{ \exp \{  {\rm i} {\vec k} ( {\vec x} - {\vec y} ) \}  }{
         a^2 {\vec k}^{\; 2} + a^2 M^2                          }
      = \frac{a}{2\pi }
        \frac{ \exp \{ - M | {\vec x} - {\vec y} | \}  }{
                  | {\vec x} - {\vec y} |               } ,
\end{eqnarray}
\begin{eqnarray}
\label{Dcor}
    D^{ijab}_{xx' }
       &=&
   \beta^{-1} \sum_n \int \frac{ d^3 p}{ (2\pi )^3 }
    D^{ijab} ( \omega_n , {\vec p} )
   e^{ {\rm i} {\vec p} ( {\vec x} - {\vec x} \; ' \; )
     + {\rm i} \omega_n  ( \tau - \tau '  )  } .
\end{eqnarray}
In order to avoid IR divergence we introduce the IR momentum cut-off
$\mu$ and replace $\Delta^{ab} ( \omega_n , 0)$ by $\Delta^{ab}
(\omega_n , \mu )$.

\section{Explicit form of the  2--loop contributions}

\setcounter{equation}{0}

The explicit expressions for the various contributions $f_\alpha$
$(\alpha = 1,2,3,4)$ to the effective potential  were obtained by
{\em (i)} performing the contraction of colour and spatial Lorentz
indices; {\em (ii)} performing the Matsubara sums by contour integral
technique; {\em (iii)} expressing the result in terms of the gluon
occupation numbers.
The first step of renormalization was
 carried out by subtracting from each
Feynman diagram the similar ones with a loop taken in the limit
$\beta \to \infty$
 in all possible ways, and neglecting all terms of the effective
potential independent of the temperature and of the background field
$v$. Below we shall see
that this  renormalization procedure removes the terms which are
linear
in
the gluon occupation numbers. Exceptions are some terms containing the
occupation numbers of the zero mode (that for gluons with the IR
 cut-off momentum $\mu$). Therefore
the `renormalized' contributions $f_\alpha$ obtained in this chapter
still contain the
 UV divergences accompanying the
IR divergences.

We analize the contributions $f_\alpha$ $(\alpha =1,2,3,4)$
and separate the pieces $f_{\alpha \; gl}$ and $f_{\alpha \; vac}$
corresponding to the gluon gas  and the vacuum, respectively:
$f_\alpha = f_{\alpha \; gl} + f_{\alpha \; vac}   $. Each of these
terms can be written as the sum $f_{\alpha \; \ldots} =
f_{\alpha \; \ldots}^0 + f_{\alpha \; \ldots}^\Delta $
where $f_{\alpha \; \ldots}^0$ and $f_{\alpha \; \ldots}^\Delta$
are the contributions for $v=0$ and the additional contributions
 due to
$v \neq 0$, respectively.  The terms depending on the occupation
number of the zero mode
 are considered as part of the
vacuum contribution.

Below we do not write out the terms of the order $\Lambda^s$ with
$s=0,1,2,3$ for each contribution $f_\alpha$ separately. However,
 we shall include all these terms in our final expression for the
  bare
effective potential in Sect. 5.

\begin{itemize}
\item {\bf Contribution $f_1$}

The `renormalized' contribution $f_1$
corresponds to  the diagrams in Fig. 2.
\begin{figure}

\vspace{5cm}

\caption{`Renormalized' contribution $f_1$ to the effective
potential.}
\end{figure}
Performing the subtractions on the r.h.s. of Eq. (\ref{f1B})
 we obtain
for the `renormalized' contribution
\begin{eqnarray}
\label{f1cor}
     f_1
     &=&
   \frac{1}{4a^4} g^2  \epsilon^{abc} \epsilon^{ade}
   \left(  {\bar D}^{ijbc}_{xx}  D^{ijde}_{xx}
  +  {\bar D}^{ijbe}_{xx} D^{jicd}_{xx}
    + {\bar D}^{iibd}_{xx} D^{jjec}_{xx}  \right)
\end{eqnarray}
with
\begin{eqnarray}
\label{Dbar}
   {\bar D}^{ijab}_{xx} &=& D^{ijab}_{xx}
      - 2 \lim_{\beta \to \infty} D^{ijab}_{xx} .
\end{eqnarray}

The calculation described in Appendix A gives the  following
expression for the renormalized glue gas contribution  for $v=0$
(see Eq. (\ref{Af1gl0})  ):
\begin{eqnarray}
\label{f1gl0}
     f_{1 \; gl}^0
    &=&
     4 g^2 \int \frac{d^3 p d^3 q}{(2\pi )^6} \frac{ n_p n_q}{pq}
     = \frac{ g^2}{ 36 \beta^4}
\end{eqnarray}
with the occupation number
$n_p = \left\lbrack \exp ( \beta p ) - 1 \right\rbrack^{-1}$
 of the gluon state with momentum ${\vec p}$.
Eq. (\ref{Af1vac0}),
leads to the vacuum contribution for $v=0$:
\begin{eqnarray}
\label{f10vac}
     f_{1 \; vac}^0
    &=&
     \frac{ g^2}{a^4} \left\lbrack
    - \frac{ \alpha^4}{ 16 \pi^4 }
    - \frac{1}{\mu a} \frac{ \alpha^2 }{ 4\pi^2}
                   \right.
               \nonumber\\
      &  &           \left.
    +  \frac{a^2}{3 \beta^2 } \frac{n_\mu}{\mu a}
     + \frac{1}{4 \mu^2 a^2 } (2n_\mu +1)(2n_\mu -1)
              \right\rbrack .
\end{eqnarray}
{}From Eq. (\ref{Af1d})
 we find
 for the additional contributions due to $v \neq 0$:
\begin{eqnarray}
\label{f1gld}
   f_{1 \;  gl}^\Delta
           &=&
    \frac{g^2}{3\pi^2 \beta^4}
   \left| \sin \frac{v}{2} \right|
 \left( \left| \sin \frac{v}{2} \right| - \frac{2\pi}{3} \right) ,
\end{eqnarray}
\begin{eqnarray}
\label{f1dvac}
   f_{1 \;  vac}^\Delta
      &=  & \frac{g^2}{3} \left\lbrack
       - \frac{ 2 \left| \sin \frac{v}{2}  \right|  }{ \pi a^2
                 \beta^2 (\mu a)  }
    \left(  N_\mu (v) + n_\mu \right)
      +  \frac{2}{3 a^2 \beta^2 (\mu a) }
    \left(  N_\mu (v) - n_\mu
    \right)        \right.
                         \nonumber\\
     &  &  \qquad  +  \left.
    \frac{1}{a^4 (\mu a)^2 }
    \left(  N_\mu (v) - n_\mu \right)
    \left(  N_\mu (v)  + 3 n_\mu      \right)
        \right\rbrack
\end{eqnarray}
in terms of the `generalized occupation number' $N_\mu (v)
 = N_{p=\mu }
(v)$ defined via Eq. (\ref{Npv}).
The terms $f_{1 \; gl}^0$ and $f_{1 \; gl}^\Delta$ contribute to the
energy density of the blackbody radiation, whereas the vacuum
contributions $f_{1 \; vac}^0$ and $f_{1 \; vac}^\Delta$ are IR
divergent and depend on the temperature.  Their dependence on the IR
momentum cut-off $\mu$ is revealed using the expansions:
\begin{eqnarray}
\label{nmuex}
        n_\mu & \approx &
        \frac{1}{\beta \mu} - \frac{1}{2} + \frac{1}{12} \beta \mu
       - \frac{1}{720} (\beta \mu )^3
       + \frac{1}{ 720 \cdot 42} (\beta \mu )^5
         + {\cal O} ( \mu^7 ) ,
\end{eqnarray}
\begin{eqnarray}
\label{Nmuex}
        N_{\mu} (v) & \approx &
      - \frac{1}{2}  + \frac{ 1}{2} A \beta \mu
      + \frac{1}{12} B (\beta \mu )^3 + {\cal O} (\mu^5 )
\end{eqnarray}
with $A = (1 - \cos v )^{-1}$, $B = A - 3A^2 $. Then we obtain:
\begin{eqnarray}
  f_{1 \; vac}^0 + f_{1 \; vac}^\Delta &=&
     \frac{2 g^2 }{ ( 2\pi )^4 } \frac{ \Lambda^6}{27}
     \left\lbrack - \frac{2}{\beta \mu^3}
                  + \frac{2}{\mu^2} A
              \right.
                        \nonumber\\
            &   &  \left.
            - \frac{\beta }{\mu } \left( \frac{1}{6}
                           + 2 A  \right)
            + \beta^2 \left(  \frac{2}{3} A - A^2  \right)
       \right\rbrack
            +  {\cal O} ( \Lambda^3 )
\end{eqnarray}
after removing the constants independent of the temperature and the
background field $v$.
Thus only the terms quadratic in the occupation numbers of the
zero modes contribute.

\item {\bf Contribution $f_2$}

Making use of the expressions (\ref{calDk})-(\ref{Dcor})
 of the free propagators, we rewrite the bare contribution
(\ref{f2B}) as follows:
\begin{eqnarray}
    f_{2B} &=& \frac{g^2}{2\pi a^4}
      e^{ - Ma }
      \left\lbrack \sigma (\mu , v )
          + 2 R ( v )
      \right\rbrack
\end{eqnarray}
where $\sigma (p, v)$ is given by Eq. (\ref{sipv})
and  from Eqs. (\ref{R0}), and (\ref{dRv}) we get:
\begin{eqnarray}
      R (v) &=& a^2 \int \frac{ d^3 p}{ (2\pi )^3 } \sigma (  p ,
v)
     \approx  \frac{a^2}{ 12 \beta^2 } + \frac{ \alpha^2 }{ 8\pi^2}
         - \frac{ a^2 }{ 2\pi \beta^2 }
 \left| \sin \frac{v}{2} \right|
            .
\end{eqnarray}
The `renormalized' contribution is
 represented by the diagrams in Fig. 3
\begin{figure}

\vspace{5cm}

\caption{`Renormalized' contribution
$f_2$ to the effective potential.}
\end{figure}
and is given by
\begin{eqnarray}
\label{f2vac}
    f_2 &=&  \frac{g^2}{2\pi a^4}
      e^{ - Ma }
      \left\lbrack a^{-1}
    \left( \sigma (\mu , v) - \sigma_\infty (\mu , v )  \right)
          + 2 \left( R (v) - R_\infty ( v ) \right)
      \right\rbrack
                \nonumber\\
    & = &  -  \frac{g^2}{2\pi a^4}
      e^{- Ma}  \left\lbrack \frac{1}{\mu a}
                 N_\mu (v)
          + \frac{a^2}{\beta^2 }
       \left(  \frac{1}{6}  - \frac{1}{\pi} \left| \sin
                \frac{v}{2} \right|  \right)
           \right\rbrack
\end{eqnarray}
with $\sigma_\infty ( \mu , v) = \lim_{\beta \to \infty}
\sigma ( \mu , v) = \sigma_\infty (\mu , 0)$ ,
 $R_\infty (v) = \lim_{\beta \to \infty} R (v) = R_\infty (0)$
given by Eqs. (\ref{sinf0}) and (\ref{Rinf0}).
Here we made use of the fact
 that the loop with the propagator ${\cal D}$ vanishes
in the limit $\beta \to \infty$ due to $M \sim \beta \to \infty$.
 The
contribution $f_2$ is IR divergent and depends on the temperature.
It can be considered as part of the
free energy density of the vacuum,
 $f_{2 \; vac} = f_2$ and $f_{2 \; gl} =0$.
 For $v=0$, i.e. $M \to \infty$ it
vanishes,  $f_{2 \; vac}^0 =0$,
 and for $v \neq 0$, i.e. finite $M $ we get $f_{2 \; vac}^\Delta =
f_2$. Using the expansion (\ref{Nmuex})  we obtain:
\begin{eqnarray}
      f_{2 \; vac}^{\Delta}  = f_2 = - \frac{ 2 g^2 }{ (2\pi )^4}
            \frac{1}{2} \Lambda^5 \beta A  + {\cal O} ( \Lambda^3 ).
\end{eqnarray}

\item {\bf Contribution $f_3$}

After Fourier transformation the bare contribution $f_{3B}$
 given by Eq.
(\ref{f3B}) takes the form:
\begin{eqnarray}
\label{f3Fou}
    f_{3B}   &=&
    -  \frac{1}{2a^8}  g^2 \epsilon^{abc} \epsilon^{gef}
    \beta^{-3} \sum_{n_1 n_2 n_3}
    \int \frac{ d^3 p d^3 q d^3 Q}{ (2\pi )^9 } \cdot
         \nonumber\\
    &  &   \cdot (2\pi )^3 \delta ( {\vec p} + {\vec q} + {\vec Q} )
    \beta \delta_{n_1 + n_2 + n_3 } \cdot
                \nonumber\\
    &   &  \cdot   a^2
  \left\lbrack
  \left( D^{kieb} (\omega_{n_1} ,  {\vec q} )
         D^{ljfc} (\omega_{n_2} , {\vec p} )
          \right.  \right.   \nonumber\\
     &  &  \left.
       + D^{kjec} (\omega_{n_1} ,  {\vec q} )
         D^{lifb} (\omega_{n_2} , {\vec p} )
  \right)
  Q^i Q^k  D^{ljga} ( \omega_{n_3} , {\vec Q} )
                            \nonumber\\
       &   &
  + \left( D^{kieb} (\omega_{n_1} ,  {\vec q} )
         D^{ljfa} (\omega_{n_2} , {\vec p} )
            \right.    \nonumber\\
     &   &   \left.
       + D^{lifb} (\omega_{n_1} ,  {\vec q} )
         D^{kjea} (\omega_{n_2} , {\vec p} )
  \right)
  p^i Q^k D^{ljgc} ( \omega_{n_3} , {\vec Q} )
                            \nonumber\\
       &   &
  + \left( D^{kjec} (\omega_{n_1} ,  {\vec q} )
         D^{ljfa} (\omega_{n_2} , {\vec p} )
                   \right.    \nonumber\\
                &  &   \left.  \left.
       + D^{ljfc} (\omega_{n_1} ,  {\vec q} )
         D^{kjea} (\omega_{n_2} , {\vec p} )
  \right)
  p^i Q^k  D^{ligb} ( \omega_{n_3} , {\vec Q} )
    \right\rbrack .
\end{eqnarray}

\begin{figure}

\vspace{4cm}

\caption{`Renormalized' contribution $f_3$ to the effective
         potential}
\end{figure}
The `renormalized' contribution of the diagrams in Fig. 4 is given by
(Appendix B):
\begin{eqnarray}
     f_3   &=&
      - \frac{1}{2} g^2  \int \frac{ d^3 p d^3 Q}{ (2\pi )^6}
           {\cal I}  ,
\end{eqnarray}
where the integrand takes the form:
\begin{eqnarray}
\label{f3intg}
   {\cal I} &=& {\cal I}_2 + {\cal I}_2 '
        + {\cal I}_1  + {\cal I}_1 '
        + {\cal I}_\beta + {\cal I}_\beta '+ {\cal I}_0 .
\end{eqnarray}
With the help of the momentum dependent functions ${\bf P} \phi_3
( {\vec q} , {\vec p} , {\vec Q} )$,   \newline
 $\Phi ( {\vec q}, {\vec p},
{\vec Q} )$, and $ \phi_0 ( {\vec q}, {\vec p}, {\vec Q} )$
defined in Appendix B by Eqs. (\ref{PFi3pQ}), (\ref{FIpQ}), and
(\ref{Fi0}), respectively,
and the `occupation numbers'
 $n_q^\pm =
\lbrack \exp ( \beta E_q^\pm )  - 1 \rbrack^{-1}$ with
 $E_q^\pm = q \pm {\rm i} v /\beta$ we find the
expressions below  for the various terms of the integrand.
(For equalities holding only up to the terms containing odd powers of
the scalar product ${\vec p} {\vec Q}$ stands $\sim$.)
 The terms quadratic in the occupation numbers give contributions to
the blackbody radiation term:
\begin{eqnarray}
\label{calI2}
   {\cal I}_2 &=&
  - (n_p^+ + n_p^- ) \frac{ 4n_Q + n_Q^+ + n_Q^-
}{4Qp}  \left\lbrack
   - \frac{1}{2} \frac{ ({\vec p}{\vec Q} \; ) {\bf P} \phi_3  }{
     ({\vec p}{\vec Q} \; )^2  - p^2 Q^2 }
   + \frac{ p^2 + Q^2 }{ (p^2 - Q^2 )^2  }  \Phi
          \right\rbrack
                   \nonumber\\
      &=&
  - (n_p^+ + n_p^- ) \frac{ 4n_Q + n_Q^+ + n_Q^- }{4Qp}
        \left(   1  + \frac{  ( {\vec p}{\vec Q} \; )^2  }{ p^2 Q^2 }
        \right)
                       \nonumber\\
       & \sim &
  - (n_p^+ + n_p^- ) \frac{ 4n_Q + n_Q^+ + n_Q^- }{4Qp}
 \frac{4}{3} ,
\end{eqnarray}
\begin{eqnarray}
\label{calI2p}
  {\cal I}_2 ' &=&
   - \frac{1}{8} (n_p^+ - n_p^- ) (n_Q^+ - n_Q^- )
       \frac{ {\bf P} \phi_3  }{  ({\vec p}{\vec Q} \; )^2 - p^2 Q^2
             }  .
\end{eqnarray}
The term ${\cal I}_1$ is linear in the occupation number at the IR
cut-off and is IR divergent:
\begin{eqnarray}
\label{calI1}
  {\cal I}_1    &=&
      \left\lbrack
     \frac{ 2n_Q + n_Q^+ + n_Q^- }{ 2Q (p^2 - Q^2 )  }
     \frac{ n_\mu^+ + n_\mu^- }{ \mu }
   + \frac{ n_Q^+ + n_Q^- }{ Q (p^2 - Q^2 ) } \frac{ n_\mu }{ \mu}
       \right\rbrack
     \Phi ( {\vec q}, {\vec p}, {\vec Q} )
                 \nonumber\\
     & &  - \frac{ 1}{ p^3 \mu }
      \left\lbrack  ( n_p^+ n_p^- + 2 n_p ) (n_\mu^+ + n_\mu^- )
            + 2 (n_p^+ + n_p^- ) n_\mu
      \right\rbrack
      \phi_0 ( {\vec q}, {\vec p}, {\vec Q} )
                     \nonumber\\
              & \sim &
           -  \frac{1}{\mu }
  \left\lbrack  (2n_Q + n_Q^+ + n_Q^- )( n_\mu^+ + n_\mu^- )
          + (n_Q^+ + n_Q^- ) n_\mu  \right\rbrack
       \cdot
                      \nonumber\\
    &  &  \cdot
    \left\lbrack  \frac{ \phi_0 ( {\vec q}, {\vec Q} , {\vec p} )
          }{  Q^3 }
      - \frac{  \Phi ( {\vec q}, {\vec p}, {\vec Q} )  }{
          2 Q (p^2 - Q^2 )        }
    \right\rbrack  ,
\end{eqnarray}
The term ${\cal I}_1 '$ vanishes for vanishing background field
identically; it is IR finite:
\begin{eqnarray}
\label{calI1p}
  {\cal I}_1 ' &=&
      -  \frac{3}{2}  \frac{ n_Q^+ - n_Q^- }{ (p^2 - Q^2 )^2  }
        (n_\mu^+ - n_\mu^- )  \Phi ( {\vec q}, {\vec p}, {\vec Q} )
              \nonumber\\
      &   &
     - \frac{2}{p^4 } (n_p^+ - n_p^- ) (n_\mu^+ - n_\mu^- )
       \phi_0 ( {\vec q}, {\vec p}, {\vec Q} )
               \nonumber\\
      &  \sim &
      - ( n_Q^+ - n_Q^- ) ( n_\mu^+ - n_\mu^- )
       \left\lbrack
        \frac{  \Phi ( {\vec q}, {\vec p}, {\vec Q} ) }{
                ( p^2 - Q^2 )^2   }
          +
         \frac{2}{ Q^4} \phi_0 ( {\vec q}, {\vec p}, {\vec Q} )
       \right\rbrack  .
              \nonumber\\
\end{eqnarray}
The terms quadratic in the occupation number at the IR cut-off are
given by:
\begin{eqnarray}
\label{calIb}
   {\cal I}_\beta &=&   \frac{1}{2p^2 \mu^2  }
      ( n_\mu^+ + n_\mu^- + 4 n_\mu )
      ( n_\mu^+ + n_\mu^- )
      \phi_0 ( {\vec q}, {\vec p}, {\vec Q} ) ,
\end{eqnarray}
and
\begin{eqnarray}
\label{calIbp}
   {\cal I}_\beta ' &=& - \frac{1}{p^4 }
      ( n_\mu^+ - n_\mu^- )^2  \phi_0 ( {\vec q}, {\vec p}, {\vec Q}
              )  .
\end{eqnarray}
The integrand ${\cal I}_\beta$ is temperature dependent and IR
divergent.
Finally, there is the IR divergent, temperature independent part of
the integrand:
\begin{eqnarray}
\label{calI0}
   {\cal I}_0 &=&
     - \frac{1}{2qpQ( p+q+Q ) }
    {\bf P} \phi_3  ( {\vec q}, {\vec p}, {\vec Q} )
      - \frac{ 3 \Phi ({\vec q}, {\vec p}, {\vec Q} )}{ 2 pQ (p+Q) }
     \left( \frac{1}{\mu } - \frac{1}{ p+Q}  \right)
                   \nonumber\\
     &   &   - \frac{3}{ p^2 \mu^2 }  \left(  1 -
      \frac{ 2\mu }{p }   \right)
      \phi_0 ( {\vec q}, {\vec p}, {\vec Q} ) .
\end{eqnarray}
The corresponding contribution to the effective potential is a part of
the energy of the perturbative vacuum.

Introducing the notation
\begin{eqnarray}
\label{f3dots}
   f_{3 \; \cdots}  &=&  - \frac{1}{2 } g^2 \int \frac{ d^3 p d^3 Q}{
                       (2\pi )^6  }   {\cal I}_{\cdots } .
\end{eqnarray}
we write
\begin{eqnarray}
  f_3 =  f_{32} + f_{32} '  + f_{31} + f_{31} '
     +   f_{3 \beta} + f_{3 \beta} '  + f_{30} .
\end{eqnarray}
Here the terms $f_{32} '$, $f_{31} '$, and $f_{3 \beta } ' $
contribute only for $v \neq 0$, whereas the other terms
can be cast into a piece for $v=0$ and the rest vanishing for $v=0$:
$ f_{32} = f_{32}^0 + f_{32}^\Delta$, $ f_{31} = f_{31}^0 +
f_{31}^\Delta$, and $ f_{3 \beta} = f_{3 \beta}^0 + f_{3\beta}^\Delta$.

The only  terms not suffering from IR divergences
are  $f_{32}$ and $f_{32} '$, the contributions to the
density of the free
energy  of the glue gas,
\begin{eqnarray}
 f_{3 \; gl} = f_{32}^0 + f_{32}^\Delta + f_{32} ' .
\end{eqnarray}
For $v=0$ we find:
\begin{eqnarray}
\label{f3gl0}
   f_{3 \; gl}^0 \equiv   f_{32}^0
     &=&   2 g^2 \int \frac{ d^3 p d^3 Q}{ (2\pi )^6 }
          \frac{ n_Q  n_p }{ Qp}
          =    \frac{g^2}{72 \beta^4 }  .
\end{eqnarray}
It is a contribution to the energy of the blackbody radiation.

For $v \neq 0$ we obtain the additional contribution to the density
of the free energy of the glue gas,
$f_{3 \; gl}^\Delta = f_{32}^\Delta + f_{32} ' $
where (see Eqs. (\ref{f32d}), (\ref{Gv}), and
 (\ref{f32pr})-(\ref{dep})
in Appendix B)
\begin{eqnarray}
\label{f32De}
  f_{32}^\Delta
       & = &
        \frac{1}{ 24\pi^4 } \frac{g^2}{\beta^4}
       \left(  \frac{ 4\pi^2}{3}  + G(v) \right) G(v)  ,
\end{eqnarray}
\begin{eqnarray}
\label{f32Pr}
   f_{32} ' & \approx &
       \frac{ 7}{ 2e^{-2}  (2\pi )^4 } \frac{g^2}{\beta^4}
       \frac{ \sin^2 v }{ (\cosh 1 - \cos v )^2 }
       \left( 1 - (1/e) \right) \left( 2 - (5/e) \right)
             \nonumber\\
\end{eqnarray}
with $G(v)$ defined by Eq. (\ref{Gv}).

 The IR divergent terms
\begin{eqnarray}
\label{f3vac}
f_{3 \; vac} &=& f_{31} + f_{3\beta} + f_{30} + f_{31} '
 + f_{3\beta} '
\end{eqnarray}
contribute to the free energy of the vacuum.
 Their dependence on the IR
cut-off $\mu $ is revealed as described at the end of Appendix B. Thus
we obtained ( $A = ( 1 - \cos v )^{-1}$ ):
\begin{eqnarray}
  f_{3 \; vac} & = &
      \frac{ 2 g^2 }{ (2\pi )^4 } \left\{
        \frac{1}{\mu^4} \frac{\Lambda^7}{\beta} \frac{ 25}{128 \cdot 7}
                \right.
                \nonumber\\
        &   &
       + \frac{1}{\mu^3} \left\lbrack
          - \Lambda^7 \frac{93}{ 128 \cdot 3 \cdot 7} A
          + \frac{ \Lambda^6}{ \beta} \frac{2431}{16 \cdot 9 \cdot
                     5 \cdot 7}
          + \frac{\Lambda^6}{12 \beta} \ln \frac{\Lambda}{\mu}
                           \right\rbrack
                  \nonumber\\
          &   &
        + \frac{1}{\mu^2} \left\lbrack
           \frac{ \Lambda^5}{\beta} \frac{33}{32 \cdot 5} \ln (\beta
                  \mu )
            - \frac{\Lambda^6}{12} A \ln \frac{\Lambda}{\mu}
                    \right.
                  \nonumber\\
          &  &
              + \Lambda^7 \beta \left( \frac{17}{16 \cdot 9 \cdot 7}
                  - \frac{1}{32 e} - \frac{5}{64 \cdot 7 } A
                 - \frac{1}{64} A^2 \right)
                   \nonumber\\
            &  &  \left.
              - \Lambda^6 \frac{ 10039}{ 64 \cdot 9 \cdot 5 \cdot 7}
                 A
              - \frac{\Lambda^5}{32 \cdot 5 \; \beta} \left(
               \frac{ 101}{3} + \frac{33}{e} - 33 A \right)
                    \right\rbrack
                 \nonumber\\
          &   &
             + \frac{1}{\mu} \left\lbrack
               \frac{\Lambda^7 \beta^2}{ 64 \cdot 3} ( A - 3A^2 )
              - \Lambda^5 \frac{561}{64 \cdot 5} \right\rbrack
             \ln (\beta \mu )
                     \nonumber\\
            &   &
              + \frac{1}{\mu} \frac{ \Lambda^6 \beta}{ 24}
                \left( \frac{1}{6} + 2A \right)
\ln \frac{\Lambda}{\mu}
                     \nonumber\\
            &   &
              + \frac{1}{\mu}  \left\lbrack
               - \frac{\Lambda^7 \beta^2}{64}
                 \left( \frac{5}{3} - \frac{4}{e}
                       - \frac{93}{9 \cdot 7} A  -
                          \frac{281}{ 4 \cdot 3 \cdot 7} A^2
                 \right)
                           \right.
                   \nonumber\\
              &  & \left.
                 + \Lambda^6 \beta \frac{2431}{ 32 \cdot 9 \cdot 5
                         \cdot 7}
                    \left(  \frac{1}{6} + 2A \right)
                 - \Lambda^5 \frac{ 1423}{ 64 \cdot 3 \cdot 5 } A
                 - \frac{ 3 \Lambda^4}{ 4 \beta}
                     \right\rbrack
                     \nonumber\\
              &  &
               + \left\lbrack \frac{\Lambda^7 \beta^3 }{ 64}
                 \left( \frac{1}{360} - \frac{A}{3} + A^2 \right)
                + \frac{ \Lambda^5 \beta}{30}
                 \left( \frac{13}{3} +
                           \frac{ 1531}{32} A \right)
                 \right\rbrack
                   \ln ( \beta \mu )
                         \nonumber\\
                &  &
              +  \left\lbrack  \frac{ \Lambda^6 \beta^2 }{48}
                       ( -A + A^2 ) +
                    \frac{\Lambda^4 }{16} A
                 \right\rbrack
                  \ln \frac{ \Lambda }{\mu }
                          \nonumber\\
                   &   &
              + \frac{ \Lambda^7 \beta^3 }{ 128}
                 \left( \frac{ 101}{ 2 \cdot 9 \cdot 5 \cdot 7 }
                    - \frac{1}{3 e}
                   + \frac{ 593}{ 2 \cdot 9 \cdot 7} A
                     - \frac{10}{e} A
                           \right.
                         \nonumber\\
                      &   &   \left.
                   - \frac{ 411}{ 2 \cdot 3 \cdot 7} A^2
                    - A^3
                   \right)
                          \nonumber\\
                       &   &
                    + \frac{ \Lambda^6 \beta^2 }{ 128 \cdot 9 \cdot 5
             \cdot 7}   \left(
                  \frac{ 10669}{2} - 5807 A \right)
                            \nonumber\\
                     &   &
                    + \Lambda^5 \beta \left(
                  - \frac{ 4851}{ 128 \cdot 9 \cdot 5}
                   + \frac{159}{ 16 \cdot 3 \cdot 5 \; e}
                  + \frac{ 1587}{ 128 \cdot 9} A
                            \right.
                           \nonumber\\
                     &  &
                    \left.
                     - \frac{77}{32 \cdot 3 \; e} A +
                        \frac{ 23}{24}  A^2   \right)
                              \nonumber\\
                       &   &
                    \left.
                     + \Lambda^4 A
     \frac{ 551}{4 \cdot 3 \cdot 5 \cdot
                      7}
                    \right\}   + {\cal O} (\Lambda^3 ) .
\end{eqnarray}
The terms independent of $\beta$ and $v$ are neglected.

\item {\bf Contribution $f_4$}

As discussed in Appendix C we obtain the expression (\ref{f4vacd})
for the contribution $f_4$
after  carrying out the subtractions shown in Fig. 5.
\begin{figure}

\vspace{4cm}

\caption{`Renormalized' contribution $f_4$ to the effective
         potential}
\end{figure}

In the perturbative regime with $v=0$, i.e. $M \to \infty$ we
get $f_4 =f_{4 \; vac }^0 =0$ from Eq. (\ref{f4vacd}).
This is because the infinitely heavy timelike gluons do not
propagate.

Following the procedure described at the end of Appendix C,
we get
\begin{eqnarray}
\label{f4Rvac}
  f_{4 \; vac}^\Delta = f_4 =
           - \frac{ 2 g^2}{ ( 2\pi )^4 }
\frac{1}{ 3 ( 6 \pi^2 )^{1/3} }
             \frac{v^2}{\beta^2} \left(
              \frac{\Lambda^5}{ \mu^3 }  + 2 \frac{
               \Lambda^4}{\mu^2 }  \right)  +
              {\cal O} ( \Lambda^3 )  .
\end{eqnarray}
Here the terms independent of $\beta$ and $v$ are neglected.

\end{itemize}

\section{Blackbody radiation}

\setcounter{equation}{0}

The complete 2--loop contribution to the density of the free
energy of
the glue gas is given by $f_{gl} = f_{gl}^0 + f_{gl}^\Delta$.

Consider at first $f_{gl}^0$ for the case $v=0$. The renormalized
diagrams 2 and 4 (Fig. 1) do not
contribute
to the glue gas as discussed in Sect. 3. The renormalized
 diagrams 1 and 3 (Fig. 1) give the  gauge
 invariant contribution to the free energy density of the glue gas
  (see Eqs. (\ref{f1gl0}) and (\ref{f3gl0}) ):
\begin{eqnarray}
\label{fgl0}
    f_{gl}^0 &=&   f_{1 \; gl}^0  + f_{3 \; gl}^0
            =
      (4 + 2) g^2 \int \frac{ d^3 p d^3 Q}{(2\pi )^6 }
           \frac{ n_Q n_p }{ Qp }
           = \frac{ g^2}{ 24 \beta^4  } .
\end{eqnarray}
This result is just the same as that one obtains by
adding the contribution of the gauge field and that of the ghosts
in the usual perturbative calculation given in \cite{Kap79}:
\begin{eqnarray}
     c_{K} g^2 \int \frac{ d^3 p d^3 Q}{(2\pi )^6 }
           \frac{ n_Q n_p }{ Qp }
\end{eqnarray}
with $ c_{K} = c_{ghost} + c_1 + c_3$, where the weights of the
ghosts, diagram 1 and diagram 3 are
   $ c_{ghost} = \frac{1}{4} \cdot 6$,
   $ c_{1} = \frac{12}{4} \cdot 6$, and
   $ c_{3} = - \frac{9}{4} \cdot 6$,  respectively, for $SU(2)$,
so that $ c_{K} =6$. The contributions of the single diagrams are
different in our case, they are gauge dependent, but the sum of them
is gauge invariant, as expected \cite{Kap79}.

For $v \neq 0$
the additional contributions to the free energy of the glue gas
are due to diagrams 1 and 3, as $f_{2 \; gl}^\Delta =
f_{4 \; gl}^\Delta =0$. According to Eqs. (\ref{f1gld}),
 (\ref{f32De}),
and (\ref{f32Pr})
we find:
\begin{eqnarray}
\label{fdgl}
   f_{gl  }^\Delta  &=&
    f_{1 \; gl }^\Delta + f_{3 \; gl }^\Delta
             \nonumber\\
     & = &
     \frac{ g^2}{ 24 \beta^4 }
     \left\lbrack  \frac{8}{\pi^2 } \left| \sin \frac{v}{2} \right|
        \left( \left| \sin \frac{v}{2} \right| - \frac{ 2\pi }{3}
        \right)
    + \frac{1}{\pi^4} \left( \frac{ 4\pi^2}{3} + G (v) \right)
          G (v)
      \right.
                     \nonumber\\
      &   &  \qquad  \left.
    + \frac{ 21 }{ 4 \pi^4 e^{-2}  } \frac{ \sin^2 v}{
     ( \cosh 1 - \cos v)^2  }  ( 1 - e^{-1} )( 2 - 5 e^{-1} )
       \right\rbrack
\end{eqnarray}
with $e$ the basis of the natural logarithm, and the function $G(v)$
given by Eq. (\ref{Gv}). This contribution vanishes for $v =0$.

The sum $f_{gl} = f_{gl}^0 + f_{gl}^\Delta$ represents the complete
contribution of the glue gas to the free energy density. This is an
 IR
finite contribution, independent of the IR cut-off momentum $\mu$.
The free energy of
the vacuum, however, depends on the background field $v$ and the IR
cut-off momentum, as well.
In order to reveal its $v$ dependence,
we have to get rid of the dependence
 on  the IR momentum cut-off.

\section{IR divergent contributions}

\setcounter{equation}{0}

Gathering  the IR divergent pieces obtained in Sect. 3 we find
the following contribution to the effective potential:
\begin{eqnarray}
\label{VIR}
     V_{IR} &=&  f_{1 \; vac}^0 + f_{1 \; vac}^\Delta
                 + f_{2 \; vac}^\Delta
        +  f_{3 \; vac}  + f_{4 \; vac}^\Delta
                 \nonumber\\
               &=& {\tilde V}_2 + V_0
\end{eqnarray}
with $ A = ( 1 - \cos v )^{-1}$ and
\begin{eqnarray}
\label{V2}
       {\tilde V}_2
            & \approx &
         \frac{2g^2}{ (2\pi )^4 }
         \left\{ \mu^{-4} \Lambda^7  \beta^{-1}  b_{41}
          \right.
                  \nonumber\\
            &   &
          \qquad  +  \mu^{-3} \left\lbrack
                     - \Lambda^7 b_{31} A
 + \Lambda^6 \beta^{-1} b_{32}
                     - \Lambda^5 \beta^{-2} b_{33} v^2
                        \right\rbrack
                  \nonumber\\
           &   &
        \qquad    + \mu^{-3} \frac{\Lambda^6}{ 12 \beta} \ln \left(
                  \Lambda /  \mu \right)
                      \nonumber\\
             &   &
       \qquad     + \mu^{-2} \left\lbrack
              \Lambda^7 \beta ( b_{21} - b_{22} A - b_{23} A^2 )
             -  \Lambda^6 b_{24} A
                      \right.
                   \nonumber\\
              &   &
        \qquad         \left.
          \qquad     + \Lambda^5 \beta^{-1} ( - b_{25} + b_{26} A )
               - \Lambda^4 \beta^{-2}  b_{27} v^2
                       \right\rbrack
                    \nonumber\\
               &  &
        \qquad       + \mu^{-2} \left\lbrack \Lambda^5 \beta^{-1}
 b_{28}
                  \ln ( \beta \mu )
                  - \Lambda^6 ( A / 12 ) \ln ( \Lambda / \mu )
                          \right.
                        \nonumber\\
               &   &
                     \left.
         \qquad    \qquad    + \Lambda^3 \beta^{-3} \Gamma_{23} (v)
                 + \Lambda \beta^{-5} \Gamma_{21} (v)
                 + \beta^{-6} \Gamma_{20} (v)
                          \right\rbrack
                       \nonumber\\
                &   &
          \qquad     + \mu^{-1}  \left\lbrack
                   \Lambda^7 \beta^2 ( - b_{11} + b_{12} A
+ b_{13} A^2
                                  )
                               \right.
                          \nonumber\\
                   &   &
            \qquad   \qquad   +  \Lambda^6 \beta ( b_{14} + b_{15} A )
                  - \Lambda^5 b_{16} A
                  - \Lambda^4 \beta^{-1} b_{17}
                              \nonumber\\
                     &   &
                             \left.
            \qquad     \qquad
 - \Lambda^3 \beta^{-2} \Gamma_{13} (v)
                      - \Lambda \beta^{-4} \Gamma_{11} (v)
                      - \beta^{-5} \Gamma_{10} (v)
                                 \right\rbrack
                            \nonumber\\
                    &   &
            \qquad
         + \mu^{-1} \left\lbrack \Lambda^7 \beta^2 b_{18}
             ( A - 3  A^2 )
                    -  \Lambda^5 ( b_{19} + b_{1 \; 10} A )
                           \right\rbrack  \ln ( \beta \mu )
                           \nonumber\\
                     &   &
            \qquad     + \mu^{-1}
               \left\lbrack \Lambda^6 ( \beta  / 144) \; ( 1 + 12 A)
                     + \Lambda^4 b_{1 \; 11} A \right\rbrack
                  \ln ( \Lambda / \mu )
                            \nonumber\\
                      &    &
           \qquad     + \Lambda^7   \beta^3 ( b_{01} + b_{02} A
           - b_{03} A^2   - b_{04} A^3   )
                + \Lambda^6 \beta^2 ( - b_{05} A + b_{06} A^2 )
                             \nonumber\\
                      &   &
           \qquad    + \Lambda^5 \beta ( - b_{07} + b_{08} A
            + b_{09} A^2 )
               + \Lambda^4  b_{0 \; 10} A
                               \nonumber\\
         &  & \qquad    + \Lambda^3 \beta^{-1} \Gamma_{03} (v)
               + \Lambda^2 \beta^{-2} \Gamma_{02} (v)
               +  \Lambda \beta^{-3}  \Gamma_{01} (v)
               +  \beta^{-4} \Gamma_{00} (v)
                               \nonumber\\
                      &   &
           \qquad   + \left\lbrack \frac{ \Lambda^7 \beta^3 }{ 64}
               \left( \frac{1}{360} - \frac{1}{3} A + A^2 \right)
               + \Lambda^5 \beta ( b_{0 \; 11} + b_{0 \; 12} A )
               - \frac{ \Lambda^3 }{ 4\beta }
                 \right\rbrack  \ln (\beta \mu )
                                \nonumber\\
                       &   &
             \qquad   + \left\lbrack
               - \frac{ \Lambda^6 \beta^2 }{ 48} ( A - A^2 )
               + \Lambda^4 b_{0 \; 13} A
               - 0.52 \Lambda^2 \beta^{-2} v^2
                   \right.
                    \nonumber\\
                 &   &  \left.  \left.
                   \qquad   \qquad
               + \frac{2.54}{\beta^4 } ( 1 - 2A )
                  \right\rbrack  \ln ( \Lambda /\mu )
               \right\}  ,
\end{eqnarray}
and
\begin{eqnarray}
\label{V0}
       V_0  & \approx &
      \frac{1}{ 4 \pi^4 ( 6 \pi^2 )^{4/3}  \beta A  }
       \left\{
         \mu^{-2} \Lambda^3 \beta^{-2} v^2  \; 9 \pi^4 \;
              \left(  \ln (\Lambda /\mu )  - 1.4 \right)
        \right.
                   \nonumber\\
             &   &
       \qquad  + \mu^{-1} \Lambda^2 \beta^{-2} v^2 \; 24 \pi^4
         + \pi^4 \left(  - 2.25 \; \Lambda^3
               + 78 \Lambda \beta^{-2}  v^2  \right)
                \ln ( \Lambda /\mu  )
             \nonumber\\
         &   &   \left.
        \qquad  - 643 \Lambda^3
          + \Lambda \beta^{-2} ( - 130 + 1022 v^2 )
       \right\}
\end{eqnarray}
 where the constants $b_{..}$ are positive numbers: \\
$b_{41} = 0.028$, $b_{31} = 0.035$, $b_{32} = 0.408$, $b_{33} =
0.086$,\\
$b_{21} = 0.0054$, $b_{22} = 0.011$, $b_{23} = 0.016$,
$b_{24} = 0.424$, $b_{25} = 0.287$, $b_{26} = 0.206$,
$b_{27} = 0.171$, $b_{28} = 0.206$, \\
$b_{11} = 0.003$, $b_{12} = 0.023$, $b_{13} = 0.052$,
$b_{14} = 0.033$, $b_{15} = 0.408$, $b_{16} = 1.48$,
$b_{17} = 0.793$, $b_{18} = 0.0052$, $b_{19} = 1.65$,
$b_{1 \; 10} = 0.667$, $b_{1 \; 11} = 0.063$, \\
$b_{01} = 0.0003$, $b_{02} = 0.008$, $b_{03} = 0.076$,
$b_{04} = 0.0078$, $b_{05} = 0.108$, $b_{06} = 0.107$,
$b_{07} = 0.60$, $b_{08} = 0.66$, $b_{09} = 0.96$,
$ b_{0 \; 10} = 1.31$, $b_{0 \; 11} = 0.145$,
$b_{0 \; 12} = 1.59$, $ b_{0 \; 13} = 0.063$, \\
and
\begin{eqnarray}
  \Gamma_{23} (v) &=& 14.35 - 0.23 A + 0.05 A^2
  - 2.79 \left| \sin \frac{v}{2} \right|  ,
          \\
  \Gamma_{21} (v) &=& -3.48 + 0.28 A - 0.054 A^2   ,
          \\
  \Gamma_{20} (v) &=& - 0.60 + 0.048 A - 0.016 A^2  ,
          \\
  \Gamma_{13} (v) &=&  17.66 - 0.23 A + 0.05 A^2
  - 2.79 \left| \sin \frac{v}{2} \right|  ,
          \\
   \Gamma_{11} (v) &=& - 0.52 + 0.28 A - 0.054 A^2 ,
          \\
   \Gamma_{10} (v) &=&  - 3.24 + 0.16 A - 0.016 A^2 ,
          \\
   \Gamma_{03} (v)  &=&  1.429 - 3.96 A - 0.182 A^2 + 0.025 A^3
              \nonumber\\
           &   &  - ( 0.233  + 1.40 A )
 \left| \sin \frac{v}{2} \right|  ,
              \\
     \Gamma_{02} (v) &=& 1.94 \; v^2 ,
               \\
    \Gamma_{01} (v) &=& - 0.064 - 0.904 A + 0.023 A^2 - 0.008 A^3 .
\end{eqnarray}

The contribution $V_0$ of the order $g^0$  arises
from the diagram $f_4$
through expansion in powers of $(\mu / M)^2 \sim \mu^2 \beta^{-1}
\Lambda^{-3} g^{-2}$. Thus terms of the order
 $g^2 (1 / g^2 ) = 1$ occur.
 The
terms of higher order in $(1/ g^2 )$ vanish as some powers of $(1
/\Lambda )$.

The bare finite temperature effective potential
$V_{eff} (\beta , v)$
as function of the constant background field $A^{03} = (a/\beta )
(v/g)$, $A^{01} = A^{02} =A^{ia} =0$  $(i=1,2,3$; $a=1,2,3)$
takes now the following form:
\begin{eqnarray}
\label{Vbare}
      V_{eff} (\beta , v)
       &= & V_W (\beta , v ) + V_H ( \beta , v)
        + \Delta V (\beta , v) + V_{IR } ,
\end{eqnarray}
where
\begin{eqnarray}
 V_W (\beta , v)  &=&
   - \frac{2 \pi^2 }{\beta^4} \left\{
    \frac{1}{45} -   \frac{1}{24}  \left\lbrack
     1 - \left( \frac{ |v|}{\pi} -1 \right)^2  \right\rbrack^2
                                          \right\}
\end{eqnarray}
 is the term obtained by Weiss \cite{Wei81},
\begin{eqnarray}
      V_H &=&
      \frac{ \alpha_0}{ a^4} \frac{ g^2}{ \sin^2 \frac{v}{2}  }
\end{eqnarray}
with $\alpha_0 = (6 \pi^2 )^{1/2}$
is the term induced by the Haar measure,
the 2--loop contribution due to the usual non-abelian
 self-interaction
is given by Eq. (\ref{fdgl}) as
\begin{eqnarray}
     \Delta V ( \beta , v ) &=&  f^\Delta_{gl} =
              \nonumber\\
   &  = &   \frac{ g^2}{ 24 \beta^4}  \left\{ 1
     + \frac{8}{\pi^2} \left| \sin \frac{v}{2} \right|
      \left( \left| \sin \frac{v}{2} \right| - \frac{ 2\pi }{3}
      \right)
     + \frac{1}{\pi^4} \left( \frac{ 4\pi^2 }{3}
         + G(v) \right)  G (v)
                 \right.
               \nonumber\\
       & & \left.
      + \frac{21}{4 \pi^4 e^{-2} }
     \frac{ \sin^2 v  }{ ( \cosh 1 - \cos v )^2  }
     \left( 1 - e^{-1} \right) \left( 2 - 5 e^{-1} \right)
             \right\} ,
\end{eqnarray}
and  the IR divergent piece is given by Eq.
(\ref{VIR}).  As the variable $A^{03}$ is compact, we have to set
forth periodically the expression of the effective potential
 given for
$v \in \lbrack 0 , 2\pi \rbrack$.

The terms periodic in $v$ arose due to the longitudinal (spatial)
gluons. As observed in \cite{Wei81} their Matsubara frequencies are
shifted by the background field, $\omega_n \pm (v/\beta )$. Having
rewritten
 the Matsubara sums over the longitudinal modes
as contour integrals on the complex energy plane, they
  are
dominated by the complex poles at $ | {\vec p} | \pm {\rm i} (v/\beta
)$. Thus the generalized occupation numbers $n_p^\pm$ and $N_p (v)$
arise and lead to  terms periodic in $v$.

Here we recover our observation made in \cite{Sai95},
that the Haar measure term $\sim + g^2
\Lambda^4 A$ of the bare effective potential is not cancelled by the
appropriate IR finite piece
$\sim + g^2 b_{0 \; 10} \Lambda^4  A$ due to the longitudinal gluons
 as $b_{0 \; 10} > 0$.
Our finding that the Haar measure term, $g^2 \Lambda^4 A$ of the bare
effective potential is not cancelled by a corresponding piece of the
contributions of longitudinal gluons is in disagreement with the
results of \cite{Boh94}. This disagreement can be the consequence
of the difference of the both approaches. We treat the quadratic
piece of the Haar measure induced potential (the last term
of Eq. (\ref{S0})) non-perturbatively, including it into the
free propagator of the field $\delta \phi$, and the first term
of Eq. (\ref{S2}) perturbatively. On the other hand both terms
are treated non-perturbatively in \cite{Boh94}.
 Below we show, however, that the Haar measure
does not influence the main qualitative features of the
renormalized effective
potential in our approximation. Thus the question of cancellation
of the Haar measure is not crucial as far as our conclusions
on the basic features of the effective potential are considered.

The bare effective potential represents  a sum of IR singular
terms, each
of them being a sum of powers of the UV cut-off.  IR divergences
appearing as powers of the dimensionless parameter
 $\Lambda /\mu $ lead
to
additional UV divergences of higher order than one would expect
 by power
counting. These are yet present although the subtractions of loop
integrals at zero temperature have been performed.

Now we remove the IR singularities. For this purpose we choose
the IR momentum cut-off
$\mu$ as a power series of the inverse of the UV cut-off (i.e. of the
lattice spacing). Actually we introduce the power series for the
dimensionless IR cut-off $\beta \mu$ in terms
 of $1 / (\beta \Lambda )$:
\begin{eqnarray}
\label{invmu}
        ( \beta \mu )^{-1} &=& \sum_{k =0}^7
 \mu_k (\beta \Lambda )^{-k}
{}.
\end{eqnarray}
The series can be terminated by $k=7$, as the terms of
 higher order will
vanish in the continuum limit when the series is inserted in the bare
effective potential.  For later use we write:
\begin{eqnarray}
\label{mus}
       ( \beta \mu )^{-s} &=& \sum_{k=0}^7 m_{sk}
 ( \beta \Lambda )^{-k}
,         \qquad (s=1,2,3,4)
\end{eqnarray}
where the coefficients are obtained by taking the powers of
(\ref{invmu}). The IR cut-off defined by  Eq. (\ref{invmu})
becomes independent of the UV cut-off for large values of $\Lambda$
and proportional to the temperature $T = 1/\beta$,
$\mu \to (1/\mu_0 ) T$.

For the sake of simplicity we neglect the logarithmically divergent
terms.
 Then the coefficients on the r.h.s. of Eq.
(\ref{invmu})
are  defined by requiring that the coefficients of $\Lambda^7$,
$\Lambda^6$, $\ldots$, $\Lambda$  vanish
in the effective potential
(\ref{Vbare}) having inserted the sum (\ref{invmu}) in it.
 These renormalization conditions
lead to highly non-linear  equations for the determination of the
coefficients $\mu_k$ $(k=0,1,\ldots ,7)$.  Assuming that our
 procedure
gives a reasonable result, i.e. the inequality  $\mu_0 \gg 1$ holds,
the coefficients $m_{sk}$ in Eq. (\ref{mus}) take the rather simple
form: $m_{s0} = \mu_0^s$, $m_{sk} = s \mu_0^{s-1} \mu_k$ for $k =
1,2,\ldots ,7$, and $s=2,3,4$.  Keeping only the terms of leading
 order
in $\mu_0$, the renormalization conditions give rise to the following
equations linearized in $\mu_k$ $(k=1,2,\ldots ,6)$:
\begin{eqnarray}
\label{la7}
       0 & = & b_{41} m_{40}  -  b_{31} A m_{30}
 + {\cal O} (\mu_0^2 ),
\end{eqnarray}
\begin{eqnarray}
\label{la6}
       0 &=& b_{41} m_{41}  + b_{32} m_{30} + {\cal O} (\mu_0^2 ) ,
\end{eqnarray}
\begin{eqnarray}
\label{la5}
       0 &=& b_{41} m_{42} - v^2 b_{33} m_{30}
 + {\cal O} (\mu_0^2 ) ,
\end{eqnarray}
\begin{eqnarray}
\label{la4}
       0 &=& b_{41} m_{43} + {\cal O} (\mu_0^2 ) ,
\end{eqnarray}
\begin{eqnarray}
\label{la3}
       0 &=& b_{41} m_{44}  + {\cal O} ( \mu_0^2 ) ,
\end{eqnarray}
\begin{eqnarray}
\label{la2}
        0 &=& b_{41} m_{45} + {\cal O} ( \mu_0^2 ) ,
\end{eqnarray}
\begin{eqnarray}
\label{la1}
        0 &=& b_{41} m_{46} + {\cal O} ( \mu_0^2 ) .
\end{eqnarray}

The approximate solution of these equations  is given by
\begin{eqnarray}
\label{solv1}
   \mu_0 = \rho_0 A, \qquad  \mu_1 = - \rho_1 ,  \qquad
   \mu_2 = v^2 \rho_2 ,
            \nonumber\\
    \mu_k \sim {\cal O} ( 1 /\mu_0 ) \approx 0
  \quad  {\mbox{  for  }}   \quad k=3, 4, \ldots , 6
\end{eqnarray}
with $\rho_0 = b_{31} / b_{41}$, $\rho_1 = b_{32} / (4 b_{41} )$,
and $\rho_2 = b_{33} / (4 b_{41} )$.
Thus the coefficients $\mu_k$ $(k \ge 3 )$ are suppressed
and can be set zero in our approximation.
 As the series terminates at $k=3$
it seems to be reasonable to choose $\mu_7 =0$.

In the continuum limit the IR momentum
cut-off
takes $\mu =0$ for $v=0$ , whereas it has its maximum value
$\mu =1.6 / \beta \approx ( \pi /2 ) T$ for $v=\pi$.
 Thus the IR momentum cut-off is much
less than the lowest non-vanishing Matsubara frequency $2\pi T$.
This seems to be a justification for considering
 the coefficient $\mu_0$
large.

Inserting the limiting value of the IR momentum cut-off in the
expression (\ref{Vbare}), we obtain for the renormalized effective
potential:
\begin{eqnarray}
      V_{R} &=&  V_W + V_{0R} + V_{2R}
\end{eqnarray}
with
\begin{eqnarray}
       V_{0R} &=& 0.097 \; v^4 \beta^{-4} ( 1 - \cos v ) ,
\end{eqnarray}
\begin{eqnarray}
      V_{2R} &=& \frac{g^2}{8\pi^4 \beta^4} \left\{
           P_0 (v) + P_1 (v)  \; A + P_2 (v) \; A^2
             - 0.40 A^3  - 0.025 A^4
                    \right.
              \nonumber\\
          &  &
      \qquad   \left.   + ( -55.5 + 15.7 v^2 )
           | \sin \frac{v}{2}  |
        +  26.3 \; \sin^2 \frac{v}{2}       \right\}
\end{eqnarray}
where
\begin{eqnarray}
       P_0 (v) &=&     25.4  - 80.5 v^2 + 0.46 v^6  ,\\
        P_1 (v)  &=&  35.5 + 1.29 v^2 , \\
        P_2 (v)  &=&  - 3.9 - 0.28 v^2 + 1.27 \sin^2 v .
\end{eqnarray}
Here we used approximately $G(v) \approx - 0.32 A -1.2$.
The purely polynomic part of the renormalized effective
 potential arises
due to the decay of the $A^{03}$ condensate into two
 transverse gluons
and a timelike fluctuation $\delta \Phi$ which then annihilate
 reproducing
 the condensate.
This is a piece of the contribution $f_4$.
 The term
$\sim g v ( A^{i1} A^{i1} + A^{i2} A^{i2} ) \delta \phi$ in Eq.
(\ref{S1}) for the cubic self-interaction is responsible for
this  subprocess.

Now we can come back to the question of the significance of the Haar
 measure as far as the renormalized effective potential is concerned.
The Haar measure term $V_H$ present in the bare effective potential
explicitly as found in \cite{Sai95} gives a contribution
of the order one to the
coefficient of $\Lambda^4$, i.e. to the r.h.s of Eq. (\ref{la4}).
Thus it means a correction of the order $1/ \mu_0^3$
to the leading term of $\mu_3$ being of the order $1/\mu_0$.
Thus we conclude that the Haar measure term does not influence
the basic features of the renormalized effective potential in
our approximation. Nevertheless it reveals itself in the details
explicitly.

The periodic terms of the renormalized effective potential
occur due to the longitudinal gluons in our approximation,
as discussed above. They
 are invariant
under the
global center transformation $ v \to ( 2\pi - v )$. The
polynomials break the global center symmetry explicitly. This
 explicit symmetry breaking
 is the consequence of
 the ansatz $\langle A^{03} \rangle \sim v$ inserted in the
tree action. It would be a way to preserve the global center
symmetry that one makes use  of the Yang-Mills action derived in
\cite{Polo90}.

Here we observed renormalizability at the order $g^2$. Although the
Haar
measure generates infinitely many vertices,
the general results for
 the non-linear sigma model \cite{Zin89} and those for
renormalizable
lattice field theories \cite{Rei88}
give the hope that the theory is renormalizable.

At the end of this section let us make some remarks on the validity of
the expression obtained for the renormalized effective potential:
$(i)$
$V_2$ tends to infinity for $v=0$, this must not be taken,
however, seriously as the expansions used for the
generalized occupation
number are only valid for $v \neq 0$.
It means that our effective potential is valid in the middle of the
 interval $v \in \lbrack 0, 2\pi \rbrack$. At the ends of the
 interval
we have to accept the perturbative results of \cite{Wei81,Kap79}.
$(ii)$ Our approach of linearizing the renormalization conditions
 and
solving them for the coefficients $\mu_k$ of the expansion of the IR
cut-off is rather crude as the coefficient $\mu_0$ becomes
 of the order
one for $v \approx \pi$.
$(iii)$ It is unclear what effect the `less dangerous' logarithmic
divergences on the effective potential have.

\section{Phase structure}

\setcounter{equation}{0}

Below we show that
the renormalized effective potential exhibits a very remarkable
feature. Namely,  it has a minimum at non-vanishing background field
 $v \neq 0$ which becomes deeper than the minimum of $V_W$ at $v=0$
describing the perturbative phase.
Furthermore, the position of the new minimum is rather close to
$v = \pi$ which corresponds to the vanishing Polyakov line
$\cos ( v/2 ) =0$, i.e. the confinement.

It is easy to see that the polynom $P_0 (v)$ has a minimum for $v
\approx 0.9 \pi$ and the periodic functions in $V_{2R}$ have their
minima at $v = \pi$. Therefore we seek the extremum of the effective
potential at $v = \pi - \varepsilon$ with $\varepsilon \ll \pi$.
 The necessary condition for having a minimum
 at $v= \pi - \varepsilon$
is given by:
\begin{eqnarray}
   0 & = &   \left.
   \frac{ \partial V_{R} }{ \partial v }
          \right|_{v=\pi - \varepsilon }
        = \left\lbrack
        \frac{ \partial V_W}{ \partial v}
     +   \frac{ \partial V_{0R} }{ \partial v}
     +   \frac{ \partial V_{2R}  }{ \partial v}
           \right\rbrack_{v=\pi - \varepsilon }
                \nonumber\\
       & = &  \beta^{-4} \left\{
         \frac{1}{3} \varepsilon +
         \left( 23.3 - 13.1 \varepsilon \right)
         + \frac{g^2 }{ 8 \pi^4 } \left( 342.4 - 6534.3 \varepsilon
                                 \right)
                     \right\}      .
\end{eqnarray}
The solution is given by
\begin{eqnarray}
      \varepsilon &=& 0.052 \left( g^2 + 53.0 \right)
          \left(  g^2 + 1.5 \right)^{-1}
\end{eqnarray}
leading to $\varepsilon  = 1.8$, $1.1$, and $0.05$ for
$g^2 =0$, $1$, and $g^2 \to \infty$, respectively.
 For strong
coupling the position of the extremum is defined by the term
$V_{2R}$.  For weak coupling the term $V_{0R}$ defines
the position of
the extremum in our approximation.
 However, then $\pi - v$ gets large and
our expression for $V_R$ is not valid any more.
It is easy to show (with accuracy of ${\cal O} (g^2 )$ ) that
 we obtain a minimum at $v \neq
0$ for any coupling, as
\begin{eqnarray}
   \left.   \frac{ \partial^2 V_R}{ \partial v^2 }
    \right|_{v=\pi - \varepsilon }  & \approx &
          \beta^{-4}  \left( 52.7 + 9.2 g^2  \right)  > 0 .
\end{eqnarray}

It is well-known that there is a minimum at $v=0$, since for weak
coupling, i.e. small values of $v$ the formula $V_W$
\cite{Wei81} modified by the
2-loop correction (\ref{fgl0}) is valid \cite{Kap79}.
Our expression  is
not valid for small values of $v$. For sufficiently strong
 coupling the
minimum find by us is in the validity range of our expression for the
effective potential. Then there should be a maximum somewhere between
the minima at $v=0$ and $v= \pi - \varepsilon$.
The value of the effective potential at $v=0$ is given by
$V_W (v=0) + g^2 / (24 \beta^4 ) = \beta^{-4} \left( - 0.44 + 0.04 g^2
\right)$. The value of the effective potential in the minimum at $v =
\pi - \varepsilon$ is given by
\begin{eqnarray}
     V_R  ( v=\pi - \varepsilon ) &=&
            \beta^{-4} \left( 18.0 - 0.26 g^2 \right)
\end{eqnarray}
for large $g^2$ (i.e. for $\varepsilon = 0.052$). Thus we conclude
 that
the minimum at $v \neq 0$ is deeper than that at $v=0$ for
 sufficiently
strong coupling: $g^2 > g_c^2$ with the critical coupling  $\alpha_c
= g_c^2 / (4\pi ) \approx 4.9$.  Using the running coupling constant
$g^2 = -G_2 / \ln (\Lambda_2^2 \beta^2 )$ with the $SU(N=2)$ scale
parameter $\Lambda_2$ and $G_2 = 4\pi (11 N /3)^{-1} $,
 the following estimate is obtained for the
critical temperature: $T_c = \Lambda_2 \exp \left( G_2 / g_c^2 \right)
\approx \Lambda_2$.

It is rather interesting that the non-trivial minimum
 of the effective
potential due to the polynomic part $P_0 (v)$ is very close
 to $v = \pi$
 corresponding to vanishing Polyakov line $W = \cos (v/2)$
\cite{Wei81,Sai95}.
As discussed above, the basic subprocess leading to this
non-perturbative minimum is the decay of the $A^{03}$
condensate in two transverse gluons and a timelike fluctuation
$\delta \phi$, which annihilate once again
into the condensate.
 This subprocess is suppressed for $v \to 0$ due to the
infinitely increasing mass $M \sim ( 1 - \cos v )^{-1/2} \to \infty$
 of the
fluctuations $\delta \phi$, but it turns out to decrease the
 free energy of the vacuum for finite values of $M$, i.e. for
$v \neq 0$. The decrease of the free energy is expected to have
a maximum for the smallest value of the mass $M$, i.e. for $v = \pi$.
In our result the small shift $\varepsilon$ of the minimum of the
 effective potential is most probably the consequence of the various
 approximations.

\section{Summary}

\setcounter{equation}{0}

In the present article we have derived the
renormalized finite temperature effective potential of the
continuum $SU(2)$ Yang-Mills theory as function of the
non-vanishing expectation value $v$ of the vector potential
component $A^{03}$, $ v = g ( \beta /a) \langle A^{03} \rangle$.
The quantum fluctuations around this mean field are taken into account
up to the order $g^2$ perturbatively. The Haar measure has been
taken into account in the definition of the partition function of
the theory in order to treat large fields correctly.
Due to the non-vanishing background field IR divergent terms
occurred in the effective potential accompanied by UV divergences.
Both  are removed by
choosing the IR momentum cut-off as a power series of the inverse of
the UV momentum cut-off. The renormalization conditions  requiring
the cancellation of UV divergences have been solved in a linearized
form. (Logarithmically divergent terms were neglected.)
In the continuum limit the IR momentum cut-off
turned out to be proportional to the temperature
and takes values smaller than the first non-vanishing Matsubara-mode.

In our approximation the main qualitative features of the
finite temperature effective potential are independent of the
Haar measure. It reveals itself only in the rather fine details.
The renormalized effective potential has a minimum at non-vanishing
background field very close to $v = \pi$.
 For sufficiently strong
coupling $g^2 > g_c^2$, i.e. below the critical temperature
$T < T_c \approx \Lambda_2$ (with the $SU(2)$ scale parameter
$\Lambda_2$)
this non-trivial minimum is deeper than the minimum
at $v=0$ describing
the perturbative phase of the theory.
 It is well-known that a
minimum at $v=\pi$ corresponds to vanishing Polyakov line, i.e.
confinement.
In our approach some important
 non-perturbative effects are already taken into account
 by the mean field,
although our treating the field fluctuations
around the mean field  perturbatively is not
really justified for strong coupling, Consequently, we cannot expect
to obtain precise information on the details of the phase transition
(the position of the non-perturbative minimum, the order of the
 phase transition, etc.)
Nevertheless, we can conclude that the main mechanism responsible
for the existence of the non-perturbative phase is the decay of the
$A^{03}$ condensate (the background field $v$) in two transverse
gluons and a timelike one.

In order to get more reliable
information on the position of the non-trivial minimum our treatment
had to be improved at least by using
the global center invariant tree level action
proposed in \cite{Polo90,Joh91}, not to break global center
symmetry explicitly at the beginning.
Then all polynomials of $v$ would be replaced by some polynomials of
periodic functions of $v$ and the non-trivial extremum would occur
exactly at $v=\pi$.

\vspace{1cm}

{\bf Acknowledgements.} One of the authors (K.S.) would like
 to express
his gratitude to J. Pol\'onyi and A. Sch\"afer for the useful
discussions and the continuous encouragement. The present work
 has been
supported by the EC project ERB-CIPA-CT92-4023, proposal 3293, and
by the Hungarian Research Fund OTKA 2192/91.

\begin{appendix}
\def\theequation{\Alph{section}.\arabic{equation}}

\section{Contribution $f_1$}

\setcounter{equation}{0}

The contribution $f_1$ is given by the expression (\ref{f1cor}).
Using Eqs. (\ref{DFou}) and (\ref{Dcor})
for the propagator we see that it is diagonal
in the spatial indices $(ij)$ for $x= x'$, $D^{ijab}_{xx} =
\frac{1}{3} \delta^{ij}
  D^{ab} $, and,  consequently,
 the matrix
   ${\bar D}^{ijab}_{xx} = \frac{1}{3}
   \delta^{ij} {\bar D}^{ab} $ defined by Eq. (\ref{Dbar}),
   as well. Here we introduced
the traces with respect of the spatial indices:
\begin{eqnarray}
     D^{ab }  &=&
\frac{a^2}{\beta} \sum_n \int \frac{d^3 p}{ (2\pi
        )^3}  \left\lbrack
       \Delta^{ab} (\omega_n , \mu )
  + 2   \Delta^{ab} (\omega_n , {\vec p} )
             \right\rbrack ,
\end{eqnarray}
and
\begin{eqnarray}
    {\bar D}^{ab }  &=&
\frac{a^2}{\beta} \sum_n \int \frac{d^3 p}{ (2\pi
        )^3}  \left\lbrack
        {\bar \Delta}^{ab} (\omega_n , \mu )
  + 2   {\bar \Delta}^{ab} (\omega_n , {\vec p} )
             \right\rbrack
\end{eqnarray}
with
\begin{eqnarray}
    \frac{a^2}{\beta} \sum_n  {\bar \Delta}^{ab}
      &=& \frac{a^2}{\beta} \sum_n  \Delta^{ab}
     - 2 \lim_{\beta \to \infty}   \frac{a^2}{\beta} \sum_n
         \Delta^{ab} .
\end{eqnarray}

Performing the sums over the colour indices we obtain the following
expression:
\begin{eqnarray}
      f_1 &=&
    \frac{g^2}{3a^4}  \left(
     {\bar D}^{33} D^{11} + {\bar D}^{11} D^{33}
     + {\bar D}^{11} D^{11}  - {\bar D}^{12} D^{12}
             \right) ,
\end{eqnarray}
where the last term vanishes identically due to $D^{12} =0$.

Further we make use of the following Matsubara sums:
\begin{eqnarray}
  \sigma ( p , 0 ) &=&   \frac{1}{\beta} \sum_n d_n ( {\vec p} )
       =  \frac{1}{2p} ( 2 n_p + 1) ,
\end{eqnarray}
\begin{eqnarray}
\label{sipv}
   \sigma (  p ,v )
 &=& \frac{1}{\beta} \sum_n d_n^\pm ( {\vec p}
)  =  \frac{1}{ 2 p } \left( 2 N_p (v) + 1 \right)
\end{eqnarray}
with the Bose-Einstein occupation number $n_p = \left( e^{\beta p} -1
\right)^{-1} $ and the `generalized occupation number'
 $N_p (v)$ defined
by
\begin{eqnarray}
\label{Npv}
       2 N_p (v) + 1 &=& \frac{ \sinh ( \beta p)  }{
        \cosh ( \beta p ) - \cos v } .
\end{eqnarray}
Let us introduce the integral:
\begin{eqnarray}
  R( v) &= & a^2  \int \frac{d^3 p}{ (2\pi )^3 } \sigma (  p , v ) .
\end{eqnarray}
For $v =0$ we find:
\begin{eqnarray}
\label{R0}
    R(0)  &=&
   a^2 \int \frac{d^3 p}{(2\pi )^3 }
    \left( \frac{n_p}{p} + \frac{1}{2p} \right)
     =  \frac{a^2}{12 \beta^2} + \frac{\alpha^2 }{8\pi^2 }
\end{eqnarray}
with $\alpha = a\Lambda = (6\pi^2 )^{1/3}$.
For $v \neq 0$ let
us write $R(v) = R(0) + \Delta R$ and $\sigma (\mu ,v) = \sigma
(\mu ,0) + \Delta \sigma$ with
\begin{eqnarray}
\label{dRv}
   \Delta R &=&
   a^2 \int \frac{ d^3 p}{(2\pi )^3 }
    \frac{ \sinh (\beta p) }{ 2p}
    \left\lbrack \frac{1}{ \cosh (\beta p) - \cos v }
      - \frac{ 1}{ \cosh (\beta p) -1 }   \right\rbrack
            \nonumber\\
     & \sim & a^2 \int \frac{d^3 p}{(2\pi )^3 } \frac{\beta p}{2p}
    \left\lbrack \frac{1}{ 1-\cos v + \beta^2 p^2 /2}
     - \frac{1}{\beta^2 p^2 /2}   \right\rbrack
             \nonumber\\
      &=&
    - \frac{a^2}{ 2\pi \beta^2 } \left| \sin \frac{v}{2} \right| ,
\end{eqnarray}
and
\begin{eqnarray}
    \Delta \sigma &=& \frac{1}{\mu} \left(
             N_\mu (v) - n_\mu  \right) .
\end{eqnarray}

For the renormalization we need  the limits
\begin{eqnarray}
\label{Rinf0}
   R_\infty (0) &=& \lim_{\beta \to \infty} R (0)
    = a^2 \int \frac{d^3 p}{(2\pi )^3 } \frac{1}{2p} =
\frac{\alpha^2}{8\pi^2} ,
\end{eqnarray}
and
\begin{eqnarray}
\label{sinf0}
  a^{-1} \sigma_\infty (\mu ,0)  &=& \lim_{\beta \to \infty}
   a^{-1}\sigma (\mu ,0)  = \frac{1}{2\mu a} .
\end{eqnarray}
Furthermore we find
 $ \Delta_\infty R = 0$, i.e. $R_\infty (v) = R_\infty (0)$
and  $\Delta_\infty \sigma = 0$.

Putting the above formulas together we obtain for $v=0$:
\begin{eqnarray}
\label{Af1gl0}
      f_{1 \; gl}^0
    &=& 4  g^2
      \int \frac{d^3 p d^3 q}{(2\pi )^6} \frac{ n_p n_q}{pq}
                    \nonumber\\
    &   &
    + 4g^2 \int \frac{ d^3 p d^3 q}{(2\pi )^6} \frac{n_p}{pq}
     - 4g^2 \int \frac{d^3 p d^3 q}{(2\pi )^6} \frac{n_p}{pq}
                \nonumber\\
     &  = & \frac{g^2}{ 36 \beta^4 } ,
\end{eqnarray}
and
\begin{eqnarray}
\label{Af1vac0}
    f_{1 \; vac}^0  &=&
     \frac{g^2}{a^4} \left\lbrack
    - a^4 \int \frac{d^3 p d^3 q}{ (2\pi )^6 } \frac{1}{pq}
                   \right.
                \nonumber\\
      &  &
      + \frac{4}{a} \left( \sigma (\mu ,0) - \sigma_\infty (\mu ,0)
        \right)  a^2 \int \frac{ d^3 p}{(2\pi )^3} \frac{n_p}{p}
                \nonumber\\
      &  &
    + \frac{2}{a} \left(  \sigma (\mu ,0) - \sigma_\infty (\mu ,0)
         - \sigma (\mu ,0)  \right) a^2 \int \frac{d^3 p}{
        (2\pi )^3 }  \frac{1}{p}
                \nonumber\\
       &  &   \left.
     + \frac{1}{a^2} \sigma (\mu ,0) \left( \sigma (\mu ,0) -
          2 \sigma_\infty (\mu ,0)  \right)
              \right\rbrack
\end{eqnarray}
leading to Eq. (\ref{f10vac}).
The terms linear in the occupation numbers cancel after
renormalization in the glue gas contribution $f_{1 \; gl}^0$.

The additional terms due to $v \neq 0$ are given by:
\begin{eqnarray}
\label{Af1d}
         f_1^\Delta  &=&
   \frac{ g^2}{3 a^4} \left\lbrack
     4 \left( \frac{1}{a} \sigma (\mu ,0 ) + 2 R(0) \right)
       \left( \frac{1}{a} \Delta \sigma + 2 \Delta R \right)
                      \right.    \nonumber\\
     &  &
   -  4 \left( \frac{1}{a} \sigma_\infty (\mu ,0 ) + 2 R_\infty (0)
 \right)
       \left( \frac{1}{a} \Delta \sigma + 2 \Delta R \right)
                         \nonumber\\
      & & + \left( \frac{1}{a} \Delta \sigma + 2 \Delta R \right)^2
                            \nonumber\\
     &  & \left.  -2
      \frac{1}{a} \Delta_\infty \sigma
       \left( a^{-1} \sigma ( \mu ,0) + 2 R(0)
       + \frac{1}{a} \Delta \sigma
 + 2 \Delta R \right)
                 \right\rbrack  .
\end{eqnarray}
Separating the contribution of the glue gas and that of the vacuum,
we arrive at Eqs. (\ref{f1gld}) and (\ref{f1dvac}).

\section{Contribution $f_3$}

\setcounter{equation}{0}

\subsection{Contracting the spatial indices}

By appropriate redefinition of the colour indices we obtain
from Eq. (\ref{f3Fou}):
\begin{eqnarray}
\label{D2Bcol}
     f_{3B}   &=&
    -  \frac{1}{2} \epsilon^{abc} \epsilon^{gef}
   \frac{g^2}{a^8} \frac{1}{\beta^3} \sum_{n_1 n_2 n_3}
    \int \frac{ d^3 q d^3 p d^3 Q}{ (2\pi )^9}  \cdot
                   \nonumber\\
   &   & \cdot   (2\pi )^3 \delta ( {\vec q} + {\vec p} + {\vec Q} )
    \beta \delta_{n_1 + n_2 + n_3 } \cdot
                \nonumber\\
    &   &  \cdot   a^2
  \left\lbrack
  \left( D^{kieb} (\omega_{n_1} ,  {\vec q} )
         D^{ljfc} (\omega_{n_2} , {\vec p} )
          \right.  \right.   \nonumber\\
     &  &  \left.
       - D^{kjeb} (\omega_{n_1} ,  {\vec q} )
         D^{lifc} (\omega_{n_2} , {\vec p} )
  \right)
  Q^i Q^k  D^{ljga} ( \omega_{n_3} , {\vec Q} )
                            \nonumber\\
       &   &
  + \left( - D^{kieb} (\omega_{n_1} ,  {\vec q} )
         D^{ljfc} (\omega_{n_2} , {\vec p} )
            \right.    \nonumber\\
     &   &   \left.
       + D^{lieb} (\omega_{n_1} ,  {\vec q} )
         D^{kjfc} (\omega_{n_2} , {\vec p} )
  \right)
  p^i Q^k  D^{ljga} ( \omega_{n_3} , {\vec Q} )
                            \nonumber\\
       &   &
  + \left( D^{kjeb} (\omega_{n_1} ,  {\vec q} )
         D^{ljfc} (\omega_{n_2} , {\vec p} )
                   \right.    \nonumber\\
                &  &   \left.  \left.
       - D^{ljeb} (\omega_{n_1} ,  {\vec q} )
         D^{kjfc} (\omega_{n_2} , {\vec p} )
  \right)
  p^i Q^k  D^{liga} ( \omega_{n_3} , {\vec Q} )
    \right\rbrack .
\end{eqnarray}

We start with  contracting  the spatial indices.
For the sake of simplicity we write:
$U = \Delta^{eb} (\omega_{n_1} , \mu )$,
${\tilde U} = \Delta^{eb} (\omega_{n_1} , {\vec q} )$,
$V = \Delta^{fc} (\omega_{n_2} , \mu )$,
${\tilde V} = \Delta^{fc} (\omega_{n_2} , {\vec p} )$,
$W = \Delta^{ga} (\omega_{n_3} , \mu )$,
${\tilde W} = \Delta^{ga} (\omega_{n_3} , {\vec Q} )$.
$U$, $V$, and $W$ correspond to the longitudinal parts, ${\tilde U}$,
${\tilde V}$, and ${\tilde W}$ to the transverse parts of the
propagators.

The expression
\begin{eqnarray}
    &  &
 \left\lbrack (U- {\tilde U} ) \frac{ q^k q^i}{ q^2} +
   {\tilde U} \delta^{ki}  \right\rbrack
 \left\lbrack (V- {\tilde V} ) \frac{ p^l p^j}{ p^2} +
   {\tilde V} \delta^{lj}  \right\rbrack
        \nonumber\\
                    &   - &
 \left\lbrack (U- {\tilde U} ) \frac{ q^k q^j}{ q^2} +
   {\tilde U} \delta^{kj}  \right\rbrack
 \left\lbrack (V- {\tilde V} ) \frac{ p^l p^i}{ p^2} +
   {\tilde V} \delta^{li}  \right\rbrack
        \nonumber\\
        &=&
  ( U- {\tilde U} ) (V - {\tilde V} )
  \frac{q^k p^l}{ q^2 p^2 } ( q^i p^j - q^j p^i )
             \nonumber\\
   &  & +  {\tilde U} ( V - {\tilde V} ) \frac{p^l}{p^2 }
      ( \delta^{ki} p^j - \delta^{kj} p^i )
     + {\tilde V} ( U - {\tilde U} ) \frac{q^k}{q^2}
      ( \delta^{lj} q^i - \delta^{li} q^j )
            \nonumber\\
    &   &   + {\tilde U} {\tilde V} ( \delta^{ki} \delta^{lj}
          - \delta^{kj} \delta^{li}  )
\end{eqnarray}
is multiplied with
\begin{eqnarray}
   Q^i Q^k \left\lbrack (W - {\tilde W} ) \frac{Q^l Q^j}{Q^2}
      + {\tilde W} \delta^{lj}   \right\rbrack
\end{eqnarray}
in the first term of the bracket.
The terms proportional to $( W - {\tilde W} )$ vanish due to the
contraction of a symmetric tensor with an antisymmetric one with
indices $ij$, and we find:
\begin{eqnarray}
  &  & (U-{\tilde U} ) ( V- {\tilde V} ) {\tilde W}
   \left( \frac{ ( {\vec q} {\vec Q} \; )^2 }{ q^2 }
   - \frac{ ( {\vec q} {\vec Q} \; ) ( {\vec q} {\vec p} \; )
      ( {\vec p} {\vec Q} \; )  }{ q^2 p^2 }
   \right)
                      \nonumber\\
   &  & +  {\tilde U} ( V - {\tilde V} ) {\tilde W}
     \left(  {\vec Q}^{\; 2} - \frac{ ( {\vec p} {\vec Q} \; )^2  }{
         p^2 }   \right)
    + (U - {\tilde U} ) {\tilde V} {\tilde W}
      2 \frac{ ( {\vec q}{\vec Q} \; )^2  }{ q^2  }
                      \nonumber\\
   &  &  + {\tilde U} {\tilde V} {\tilde W} 2 {\vec Q}^{\; 2} .
\end{eqnarray}
Proceeding similarly in the second and third terms
 we find for the bracket in Eq.
(\ref{D2Bcol}):
\begin{eqnarray}
 {\tilde U} {\tilde V} {\tilde W}
        \phi_3 ({\vec q}, {\vec p},{\vec Q})
   + {\tilde U} V {\tilde W}
        \phi_1 ({\vec q}, {\vec p},{\vec Q})
   +U {\tilde V} {\tilde W}
        \phi_2 ({\vec q}, {\vec p},{\vec Q})
     +
    U V {\tilde W}
        \phi_0 ({\vec q}, {\vec p},{\vec Q})
               \nonumber\\
\end{eqnarray}
with
\begin{eqnarray}
\label{Fi3}
        \phi_3 ({\vec q}, {\vec p},{\vec Q})
     &=&
   {\vec Q}^{\; 2} - 2 ({\vec p}{\vec Q} \; )
   + \frac{ ({\vec p}{\vec Q} \; )^2}{ p^2}
   - \frac{ ({\vec q}{\vec Q} \; )^2}{q^2}
                   \nonumber\\
     & &
      + 2 \frac{ ({\vec q}{\vec p} \; )
                 ({\vec q}{\vec Q} \; )  }{ q^2 }
      -  \frac{ ({\vec q}{\vec p} \; )({\vec p}{\vec Q} \; )
                 ({\vec q}{\vec Q} \; )  }{ q^2 p^2 } ,
\end{eqnarray}
\begin{eqnarray}
\label{Fi2}
        \phi_2 ({\vec q}, {\vec p},{\vec Q})
      &=&
    \frac{ ({\vec q}{\vec Q} \; )^2}{q^2}
      - 2 \frac{ ({\vec q}{\vec p} \; )
                 ({\vec q}{\vec Q} \; )  }{ q^2 }
      +  \frac{ ({\vec q}{\vec p} \; )({\vec p}{\vec Q} \; )
                 ({\vec q}{\vec Q} \; )  }{ q^2 p^2 } ,
\end{eqnarray}
\begin{eqnarray}
\label{Fi1}
        \phi_1 ({\vec q}, {\vec p},{\vec Q})
      &=&
          {\vec Q}^{\; 2}
   - \frac{ ({\vec p}{\vec Q} \; )^2}{ p^2}
   - \frac{ ({\vec q}{\vec Q} \; )^2}{q^2}
      +  \frac{ ({\vec q}{\vec p} \; )({\vec p}{\vec Q} \; )
                 ({\vec q}{\vec Q} \; )  }{ q^2 p^2 } ,
\end{eqnarray}
\begin{eqnarray}
\label{Fi0}
        \phi_0 ({\vec q}, {\vec p},{\vec Q})
     &=&
    \frac{ ({\vec q}{\vec Q} \; )^2}{q^2}
      -  \frac{ ({\vec q}{\vec p} \; )({\vec p}{\vec Q} \; )
                 ({\vec q}{\vec Q} \; )  }{ q^2 p^2 } .
\end{eqnarray}

The contribution $f_{3B}$ takes now the form:
\begin{eqnarray}
   f_{3B}     &=&
   - \frac{1}{2} g^2 \epsilon^{abc} \epsilon^{gef}
  \frac{a^2}{a^8} \frac{1}{\beta^3}
   \sum_{n_1 n_2 n_3 } \int \frac{d^3 q d^3 p d^3 Q}{ (2\pi )^6 }
  \delta ({\vec q} + {\vec p} + {\vec Q} \; )
   \beta \delta_{n_1 + n_2 + n_3 }
                \cdot
             \nonumber\\
    &  &  \cdot
   \left\lbrack
   \Delta^{eb} ( \omega_{n_1} , {\vec q} )
   \Delta^{fc} ( \omega_{n_2} , {\vec p} )
   \Delta^{ga} ( \omega_{n_3} , {\vec Q} )
   \phi_3 ( {\vec q}, {\vec p}, {\vec Q} )
                \right.
                      \nonumber\\
   &   &  +   \Delta^{eb} ( \omega_{n_1} , \mu )
        \Delta^{fc} ( \omega_{n_2} , {\vec p} )
        \Delta^{ga} ( \omega_{n_3} , {\vec Q} )
   \phi_2 ( {\vec q}, {\vec p}, {\vec Q} )
                   \nonumber\\
   &   &  +   \Delta^{eb} ( \omega_{n_1} , {\vec q} )
        \Delta^{fc} ( \omega_{n_2} , \mu )
        \Delta^{ga} ( \omega_{n_3} , {\vec Q} )
   \phi_1 ( {\vec q}, {\vec p}, {\vec Q} )
                    \nonumber\\
   &   &      \left.
 +   \Delta^{eb} ( \omega_{n_1} , \mu  )
     \Delta^{fc} ( \omega_{n_2} , \mu )
     \Delta^{ga} ( \omega_{n_3} , Q )
   \phi_0 ( {\vec q}, {\vec p}, {\vec Q} )
             \right\rbrack     .
\end{eqnarray}

\subsection{Matsubara sums}

Let us now introduce the sum:
\begin{eqnarray}
  S (  q,  p,  Q )
  &=&
 \frac{1}{\beta^3} \sum_{n_1 n_2 n_3} \beta \delta_{n_1 + n_2 + n_3 }
 \epsilon^{abc} \epsilon^{gef}
   \Delta^{eb} ( \omega_{n_1} , {\vec q} )
   \Delta^{fc} ( \omega_{n_2} , {\vec p} )
   \Delta^{ga} ( \omega_{n_3} , {\vec Q} )
               \nonumber\\
   &=& \sum_{ permutations \; of  \; ( q  p Q )}
  s_B (  q,  p, Q )
\end{eqnarray}
with
\begin{eqnarray}
   s_B (  q,  p,  Q ) &=&
   \frac{1}{\beta^3} \sum_{n_1 n_2 n_3} \beta
 \delta_{n_1 + n_2 + n_3
}
      d_{n_1}^- ( {\vec q} ) d_{n_2}^+ ({\vec p} )  d_{n_3} ( {\vec
Q} )  .
\end{eqnarray}
With the appropriate change of the momentum variables we can write:
\begin{eqnarray}
\label{f3Bs}
        f_{3B} &=&
   - \frac{1}{2}  \frac{g^2}{a^4} a^2
  \int \frac{ d^3 q d^3 p d^3 Q}{ (2\pi )^6}
     \delta ( {\vec q} + {\vec p} + {\vec Q} \; )
           \cdot      \nonumber\\
     &   &  \cdot
    \left\lbrack
     S (  q,  p,  Q ) \phi_3 ({\vec q}, {\vec p},
                         {\vec Q} )
             \right.
                      \nonumber\\
   &  & +   S ( \mu ,  p,  Q ) \phi_2 ({\vec q}, {\vec p},
                         {\vec Q} )
    +   S (  q, \mu ,  Q ) \phi_1 ({\vec q}, {\vec p},
                         {\vec Q} )
                      \nonumber\\
   &   &    \left.
    +   S ( \mu , \mu ,  Q ) \phi_0 ({\vec q}, {\vec p},
                         {\vec Q} )
      \right\rbrack
                      \nonumber\\
   &=&
   - \frac{1}{2}  \frac{g^2}{a^4} a^2
  \int \frac{ d^3 q d^3 p d^3 Q}{ (2\pi )^6}
     \delta ( {\vec q} + {\vec p} + {\vec Q} \; )
           \cdot      \nonumber\\
     &   &  \cdot
    \left\lbrack
     s_B (  q,  p,  Q ) {\bf P} \phi_3 ({\vec q},
           {\vec p},  {\vec Q} )
             \right.
                      \nonumber\\
   &  & +   \left(
  s_B ( \mu , {\vec p}, {\vec Q} )  +
  s_B (  {\vec p}, \mu , {\vec Q} )  +
  s_B (  {\vec Q},  {\vec p} , \mu )  \right)
      \Phi ({\vec q}, {\vec p}, {\vec Q} )
                     \nonumber\\
         &  &  \left.
    + 2  \left( s_B ( \mu , \mu , {\vec Q} ) + s_B ( \mu ,  {\vec Q},
                 \mu )
                +  s_B ( {\vec Q}, \mu  ,\mu  )       \right)
      \phi_0 ({\vec q}, {\vec p},  {\vec Q} )
      \right\rbrack  ,
                      \nonumber\\
\end{eqnarray}
with
\begin{eqnarray}
\label{PFi3pQ}
  {\bf P} \phi_3 &=&
   8 ( p^2 + Q^2 ) - 2 {\vec p}{\vec Q}
  - 6 ({\vec p}{\vec Q} \; )^2 \frac{p^2 + Q^2 }{ p^2 Q^2 }
     - 2 \frac{ ({\vec p}{\vec Q} \; )^3 }{ p^2 Q^2 }
                    \nonumber\\
     &   & +   2
    \left( 2 - {\vec p}{\vec Q}  \frac{ p^2 + Q^2 }{ p^2 Q^2 }
    \right)   \cdot
                         \nonumber\\
     &   &   \cdot
   \frac{ p^2 Q^2 (p^2 + Q^2 ) + (p^4 + Q^4 ) ({\vec p}{\vec Q} \; )
   - (p^2 + Q^2 ) ({\vec p}{\vec Q} \; )^2 -
   2 ({\vec p}{\vec Q} \; )^3      }{
     (p^2 + Q^2 )^2  - 4 ({\vec p}{\vec Q} \; )^2     }  ,
                 \nonumber\\
\end{eqnarray}
\begin{eqnarray}
\label{FIpQ}
   \Phi &=&
  ( p^2 + Q^2 ) \left(  1 - \frac{ ({\vec p}{\vec Q} \; )^2  }{
          p^2 Q^2 }    \right)
                       \nonumber\\
    &   &  + 2 \left( \frac{ {\vec p}{\vec Q} }{ p^2 Q^2 }
      (p^2  + Q^2 ) - 2  \right)  \cdot
                      \nonumber\\
     &   &   \cdot
    \frac{ (p^2 + Q^2 ) p^2 Q^2 +
      (p^4 + Q^4 ) {\vec p}{\vec Q}
    - (p^2 + Q^2 ) ({\vec p}{\vec Q} \; )^2 - 2 ({\vec p}{\vec Q} \;
         )^3   }{
          (p^2 + Q^2 )^2 - 4 ({\vec p}{\vec Q} \; )^2     }  ,
                      \nonumber\\
\end{eqnarray}
where we inserted
 ${\vec q} = - {\vec p}- {\vec Q}$.

 We write
for the Kronecker-delta:
\begin{eqnarray}
   \beta \delta_{n_1 + n_2 + n_3 } &=&
 \frac{ e^{ \beta ( q^0 + p^0 + Q^0 ) } -1  }{
   q^0 + p^0 + Q^0 }  \equiv I (q^0 , p^0 , Q^0 ) ,
\end{eqnarray}
and perform the sum $s_B$ over the Matsubara  frequencies
 by the contour
integral technique of finite temperature field theory \cite{Kapbook}:
\begin{eqnarray}
   s_B ( q,  p,  Q )
    &=& -  \frac{ 1}{ 8 qpQ}
h (  q,  p,  Q )
\end{eqnarray}
with
\begin{eqnarray}
   h ( q , p , Q ) &=&
  \left\lbrack \begin{array}{l}
     n_q^-  \\ n_q^+ +1  \end{array}  \right.
  \left\lbrack \begin{array}{l}
     n_p^+  \\ n_p^- +1  \end{array}  \right.
  \left\lbrack \begin{array}{l}
     n_Q  \\ n_Q +1  \end{array}  \right.
      I ( \pm E_q^{\mp} , \pm E_p^\pm , \pm E_Q ) ,
                   \nonumber\\
\end{eqnarray}
and $E_q^\pm = q \pm {\rm i} v /\beta$.

We express the sum $h$ through the occupation numbers. Then
we separate the vacuum term $h_0$, and the terms $h_1$ and
$h_2$ linear and quadratic in the occupation numbers, respectively,
$   h = h_0 + h_1 + h_2   $
with
\begin{eqnarray}
\label{h0}
   h_0 (q,p,Q) &=&  \frac{2}{q+p+Q} ,
\end{eqnarray}
\begin{eqnarray}
\label{h1}
    h_1 (q,p,Q) &=& 4 n_Q \frac{q+p}{ (q+p)^2 - Q^2 }
      + 2 ( n_p^+ + n_p^- ) \frac{ q+Q}{ (q+Q)^2 -p^2 }
                   \nonumber\\
    &   &  + 2 (n_q^+ + n_q^- ) \frac{ p+Q}{ (p+Q)^2 -q^2 }  ,
\end{eqnarray}
and
\begin{eqnarray}
\label{h2}
   h_2 (q,p,Q) &=&
 4 n_Q (n_p^+ + n_p^- ) q \frac{ q^2 -p^2 - Q^2 }{
      q^4 + p^4 + Q^4 - 2 q^2 p^2 - 2 q^2 Q^2 - 2 p^2 Q^2 }
            \nonumber\\
  &  & +
 4 n_Q (n_q^+ + n_q^- ) p \frac{ p^2 -q^2 - Q^2 }{
      q^4 + p^4 + Q^4 - 2 q^2 p^2 - 2 q^2 Q^2 - 2 p^2 Q^2 }
            \nonumber\\
   &  &  +
 2  \frac{
  (n_q^+ + n_q^- ) ( n_p^+ + n_p^- ) Q ( Q^2 - q^2 - p^2 ) }{
      q^4 + p^4 + Q^4 - 2 q^2 p^2 - 2 q^2 Q^2 - 2 p^2 Q^2 }
              \nonumber\\
    &  &  -
  4 \frac{
  ( n_q^+ - n_q^- )( n_p^+ - n_p^- )  Qqp
  }{
      q^4 + p^4 + Q^4 - 2 q^2 p^2 - 2 q^2 Q^2 - 2 p^2 Q^2 }   .
\end{eqnarray}
These expressions are symmetric in the momenta for
vanishing background field $v=0$, as expected.

\subsection{Subtraction of loop-integrals at zero temperature}

The `renormalized' contribution $f_3$ is given
in terms of the diagrams in Fig. 4.
As the momentum dependent functions $\phi_3$, $\phi_0$,
and $\Phi$ are independent of the
temperature, we can perform the subtractions on the sum $s_B$
directly.
 We have
to subtract the limiting values of   $s_B$ when two of the occupation
numbers with all possible choices are taken in the limit
 $\beta \to \infty$.

  If none of the momentum
variables is equal to zero, we get:
\begin{eqnarray}
   h_\infty^{(pq)} ( {\vec q}, {\vec p}, {\vec Q} )
      &=&
  \frac{2}{q+p+Q}  +  4 \frac{q+p}{ (q+p)^2 - Q^2  }  n_Q ,
            \nonumber\\
   h_\infty^{(qQ)} ( {\vec q}, {\vec p}, {\vec Q} )
      &=&
  \frac{2}{q+p+Q}  +  2 \frac{q+Q}{ (q+Q)^2 - p^2  }  (n_p^+ + n_p^-
             ),
               \nonumber\\
   h_\infty^{(pQ)} ( {\vec q}, {\vec p}, {\vec Q} )
      &=&
  \frac{2}{q+p+Q}  +  2 \frac{p+Q}{ (p+Q)^2 - q^2  }  (n_q^+ + n_q^-
             )
\end{eqnarray}
and for the renormalized sum $s (q,p,Q)$:
\begin{eqnarray}
\label{sren}
    s (q,p,Q )  &= &
      s_B (q, p, Q) + \frac{1}{8pqQ}
      \left\lbrack
   h_\infty^{(pq)} ( {\vec q}, {\vec p}, {\vec Q} )
  +   h_\infty^{(qQ)} ( {\vec q}, {\vec p}, {\vec Q} )
  +    h_\infty^{(pQ)} ( {\vec q}, {\vec p}, {\vec Q} )
     \right\rbrack
                \nonumber\\
           & \sim &
       -  \frac{n_Q (n_p^+ + n_p^- ) }{ pQ} \frac{ q^2 -p^2 -Q^2 }{
   q^4 + p^4 + Q^4 - 2 q^2 p^2 - 2 q^2 Q^2 - 2 p^2 Q^2 }
                \nonumber\\
    &   &
    - \frac{ (n_Q^+ + n_Q^- )( n_p^+ + n_p^- ) }{
       4 p Q}  \frac{ q^2 - p^2 -Q^2 }{
   q^4 + p^4 + Q^4 - 2 q^2 p^2 - 2 q^2 Q^2 - 2 p^2 Q^2 }
                 \nonumber\\
    &  &
   + \frac{ (n_p^+ - n_p^- )( n_Q^+ - n_Q^- ) }{ 2Qp}
        \frac{ Qp}{
   q^4 + p^4 + Q^4 - 2 q^2 p^2 - 2 q^2 Q^2 - 2 p^2 Q^2 }
                      \nonumber\\
    &  &
   + \frac{1}{ 2qpQ (q+p+Q) }  .
\end{eqnarray}
(The $\sim$ sign means that $s (q,p,Q)$ can be replaced in the
integrand by the r.h.s. of Eq. (\ref{sren}).)
Inserting ${\vec q} = - {\vec p} - {\vec Q}$ we obtain:
\begin{eqnarray}
   s ( q, p, Q) &=&
    - \frac{1}{8Qp}
       (4 n_Q + n_Q^+ + n_Q^- ) (n_p^+ + n_p^- )
     \frac{ {\vec p}{\vec Q}  }{  ({\vec p}{\vec Q} \; )^2  - p^2 Q^2
          }
                      \nonumber\\
    &   &  + \frac{1}{8}
    ( n_p^+ - n_p^- ) (n_Q^+ - n_Q^- )  \frac{ 1}{
      ( {\vec p}{\vec Q} \; )^2   - p^2 Q^2      }
                         \nonumber\\
     &   &
     + \frac{1}{ 2qpQ( q+p+Q ) }  .
\end{eqnarray}
The terms linear in the occupation numbers cancel due to
renormalization, similarly as in the perturbative calculation without
background field \cite{Kap79}.

If one or two of the momentum variables take the value of the IR
cut-off $\mu$,  we make an expansion in its powers.
The UV renormalization is performed
by subtracting the corresponding limiting values
with $\beta \to \infty$
for fixed finite $\mu$.  It removes the terms linear in the
occupation numbers.  Thus we obtain:
\begin{eqnarray}
\lefteqn{
  s (p, \mu , Q) + s (\mu , p , Q) + s (Q, p, \mu )
          =   }
                   \nonumber\\
    & = &
     \frac{ (4n_Q + n_Q^+ + n_Q^- )( n_p^+ + n_p^- )  }{ 4Qp }
    \frac{ p^2 + Q^2 }{ (p^2 - Q^2 )^2  }
                   \nonumber\\
   &  &  - \frac{ n_Q (n_\mu^+ + n_\mu^- ) }{ Q (p^2 - Q^2 )  \mu }
      - \frac{ n_\mu ( n_Q^+ + n_Q^- ) }{ 2Q (p^2 - Q^2 ) \mu }
                   \nonumber\\
    &  &
    + \frac{ n_\mu ( n_p^+ + n_p^- ) }{ 2p (p^2 - Q^2 ) \mu }
    + \frac{ (n_p^+ + n_p^- )( n_\mu^+ + n_\mu^- ) }{
             2p\mu (p^2 - Q^2 )                    }
                          \nonumber\\
     &  &
   + \frac{  \left\lbrack
2 (n_p^+ - n_p^- ) + n_Q^+ - n_Q^-    \right\rbrack
     (  n_\mu^+ - n_\mu^- )
      }{ 2 (p^2 - Q^2 )^2  }
                           \nonumber\\
     &  &
   +  \frac{ 3}{ 2 pQ (p+Q)  }  \left( \frac{1}{\mu }
                - \frac{ 1}{ p+Q}    \right)  ,
\end{eqnarray}
\begin{eqnarray}
\lefteqn{
    s (\mu , \mu , p) + s (\mu , p , \mu ) + s ( p, \mu ,
\mu )         =
         }    \nonumber\\
     &=&
 + \frac{1}{2p^3 \mu }    \left\lbrack
 ( 2n_p + n_p^+ + n_p^- )( n_\mu^+ + n_\mu^- )
         + (n_p^+ + n_p^- ) n_\mu  \right\rbrack
                   \nonumber\\
    &  & - \frac{1}{4p^2 \mu^2}
   ( n_\mu^+ + n_\mu^- + 4n_\mu )( n_\mu^+ + n_\mu^- )
                      \nonumber\\
      &  &
           +  \frac{1}{2p^4}
    \left\lbrack
     (n_\mu^+ - n_\mu^- )^2 + 2 (n_\mu^+ - n_\mu^- )
      (n_p^+ - n_p^- )   \right\rbrack
                      \nonumber\\
     &  &  + \frac{3}{2 p^2 \mu^2  }
     \left(  1  - \frac{ 2\mu }{p }   \right)  .
\end{eqnarray}

The `renormalized' contribution $f_3$ can now be rewritten as
\begin{eqnarray}
             f_3  &= &
      - \frac{g^2}{2}  \int \frac{ d^3 p d^3 Q}{ (2\pi )^6}
           {\cal I}  ,
\end{eqnarray}
with the integrand
\begin{eqnarray}
   {\cal I} &=& - s (q,p,Q) {\bf P} \phi_3 ({\vec q}, {\vec p},
{\vec Q} )
                \nonumber\\
     &   &   - \left(  s (\mu,p,Q)  + s (p, \mu , Q)
     + s ( Q, p ,\mu )  \right)  \Phi ({\vec q}, {\vec p}, {\vec
                   Q})
                      \nonumber\\
        &  &
     - 2 \left( s (\mu, \mu, Q) + s (\mu, Q, \mu )
             + s ( Q, \mu , \mu )  \right)
      \phi_0 ( {\vec q}, {\vec p}, {\vec Q} )
\end{eqnarray}
and ${\vec q} = - {\vec p} - {\vec Q}$.
This can be cast in the terms of the order  $n_\mu^0$, $n_\mu$, and
$n_\mu^2$ depending on the occupation numbers,
 and the term independent
of the occupation numbers according to Eq. (\ref{f3intg}).

\subsection{Momentum integrals}

Here is appropriate to make some remarks on the calculation of the
momentum integrals.

The momentum dependent functions multiplying the Matsubara sums
 in the
integrands (\ref{calI2})-(\ref{calI0})
 can be replaced  by more simple
expressions.
 They are obtained by separating the terms of the
integrands with odd powers of the scalar product
 $( {\vec p}  {\vec Q} \; )$
 and leaving them out because they do not
contribute to the integrals. Equations valid only up to those terms
are denoted by $\sim $.  Thus the following expressions are found:
\small
\begin{eqnarray}
  \phi_0 ( {\vec q}, {\vec Q}, {\vec p} )
    & \sim &
   \frac{ p^2 Q^2 - ({\vec p}{\vec Q} \; )^2 }{ 2Q^2 }
    + \frac{ p^4 - Q^4 }{ 8Q^2 }
 -   \frac{ (p^2 - Q^2 )^2 (p^4 - Q^4 )  }{
    8 Q^2  \left\lbrack
    ( p^2 + Q^2 )^2 - 4 ( {\vec p}{\vec Q} \; )^2  \right\rbrack } ,
                      \nonumber\\
\end{eqnarray}
\begin{eqnarray}
  \frac{ ({\vec p}{\vec Q} \; ) {\bf P} \phi_3 ({\vec q}, {\vec p},
{\vec Q} )   }{   ({\vec p}{\vec Q} \; )^2  - p^2 Q^2 }
                &  \sim &
      - 2 \frac{ ({\vec p}{\vec Q} \; )^2 }{ p^2 Q^2 }
      - \frac{1}{2} \left\lbrack 4
        +  \frac{ (p^2 + Q^2 )^2  }{ p^2 Q^2 }  \right\rbrack
                \nonumber\\
     &   &
    + \frac{1}{2} \left\lbrack 4 + \frac{ (p^2 + Q^2 )^2 }{ p^2 Q^2 }
                   \right\rbrack
     \frac{ ( p^2 + Q^2 )^2 }{ ( p^2 + Q^2 )^2  - 4 ({\vec p}{\vec Q}
                             \; )^2    }      ,
\end{eqnarray}
\begin{eqnarray}
   \frac{ {\bf P} \phi_3 }{ ( {\vec p}{\vec Q} \; )^2 -p^2 Q^2 }
      & \sim &
    - \frac{ p^2 + Q^2 }{ p^2 Q^2 }
    \left\lbrack 7 +
     \frac{ p^4 + Q^4 + 6p^2 Q^2 }{ 2 (p^2 + Q^2 ) p Q }
    \ln \left| \frac{ p+Q}{p-Q} \right|
    \right\rbrack
             \nonumber\\
\end{eqnarray}
(obtained after performing the angle integration) ,
\begin{eqnarray}
  \frac{ 2 \Phi ( {\vec q}, {\vec p}, {\vec Q} )  }{ p^2 - Q^2 }
      &  \sim &
  2 ( p^2 + Q^2 ) (p^2 - Q^2 ) \left\{
     - \frac{1}{4 p^2 Q^2 }
            \right.
                          \nonumber\\
     &   &  \left.
   +  \left\lbrack 1 + \frac{ (p^2 + Q^2 )^2 }{ 4 p^2 Q^2 }
       \right\rbrack
     \frac{1}{  (p^2 + Q^2 )^2 - 4 ({\vec p}{\vec Q} \; )^2  }
                  \right\} .
\end{eqnarray}
\normalsize

There is a remarkable cancellation of several terms
in the integrand of ${\cal I}_2 $
(see Eq. (\ref{calI2}) ):
\small
\begin{eqnarray}
  &  &
  \left\lbrack \frac{ p^2 + Q^2 }{ (p^2 - Q^2 )^2 }
     \Phi ( {\vec q}, {\vec p}, {\vec Q} )
    -  \frac{1}{2} \frac{ ( {\vec p}{\vec Q} \; ) {\bf P}
      \phi_3 ({\vec q}, {\vec p}, {\vec Q} ) }{
      ( {\vec p}{\vec Q} \; )^2  - p^2 Q^2   }
   \right\rbrack
                \nonumber\\
   &  \sim &
   \frac{1}{2} \left\{
   2 (p^2 + Q^2 )^2
   \left\lbrack
   - \frac{1}{4p^2 Q^2 }  +
   \left( 1 + \frac{ (p^2 + Q^2 )^2 }{ 4p^2 Q^2 }  \right)
   \frac{1}{  ( p^2 + Q^2 )^2 - 4 ({\vec p}{\vec Q} \; )^2  }
   \right\rbrack
                     \right.
          \nonumber\\
    &  &
    + 2 \frac{ ({\vec p}{\vec Q} \; )^2  }{  p^2 Q^2 }
   +  2  \left(   1  + \frac{ (p^2 + Q^2 )^2 }{ 4p^2 Q^2 } \right)
               \nonumber\\
     &   &  \left.
    - 2 \left(  1  + \frac{ (p^2 + Q^2 )^2 }{ 4 p^2 Q^2 } \right)
     \frac{  (p^2 + Q^2 )^2  }{
     (p^2 + Q^2 )^2 - 4 ({\vec p}{\vec Q} \; )^2 }
             \right\}
                     \nonumber\\
     & = &
     1 + \frac{ ({\vec p}{\vec Q} \; )^2 }{ p^2 Q^2 }
            \sim \frac{4}{3}  .
\end{eqnarray}
\normalsize
The last equation is obtained by performing the angle integration.

For evaluating the  glue gas
contribution
$f_{32}^\Delta$ the following
approximation is used:
\begin{eqnarray}
  n_p^+ + n_p^- - 2n_p &=&
        \frac{1}{ e^{\beta p} e^{ {\rm i} v} - 1  }
      + \frac{1}{ e^{\beta p} e^{ - {\rm i} v} - 1  }
      - \frac{2}{ e^{\beta p}  - 1  }
            \nonumber\\
     &  =&
    \frac{ 2 \left( e^{\beta p} \cos v - 1 \right)  }{
       e^{ 2\beta p}  - 2 e^{\beta p} \cos v +1      }
      -  \frac{2}{ e^{\beta p} - 1 }
        \nonumber\\
      & \approx &
        \left\{  \begin{array}{lll}
        0  & {\mbox{for}} & p > 1/\beta \\
        \nu ( p ) &  {\mbox{for}} & p < 1/\beta
             \end{array}      \right.
\end{eqnarray}
where
\begin{eqnarray}
     \nu (p) &=&
      \frac{2}{ e^{\beta p}  }
     \left\lbrack
     \frac{ \cos v - e^{ - \beta p}  }{
     1 - 2 e^{ -\beta p} \cos v + e^{ - 2\beta p}   }
    -   \frac{1}{ 1 - e^{ - \beta p}  }
     \right\rbrack .
\end{eqnarray}
Then we obtain from Eqs. (\ref{calI2}), and (\ref{f3dots}):
\begin{eqnarray}
\label{f32d}
         f_{32}^\Delta
        & \approx &
   2 \frac{ g^2}{ (2\pi )^4 } \int_\mu^{1/\beta}
       dp  \int_\mu^\Lambda dQ \frac{pQ}{3}
     \left\lbrack 8 n_Q \nu (p)  + \nu (p) \nu (Q)
      \right\rbrack .
\end{eqnarray}
We use
 the saddle point approximation and replace the
expression $p \nu (p)$ by its value taken at $p_0 = 1 /\beta$:
\begin{eqnarray}
\label{Gv}
   G (v) & \equiv &
     \beta p_0 \nu ( p_0 ) =
      \frac{2}{ e  }
     \left\lbrack
     \frac{ \cos v - (1/ e ) }{
     1 - 2 (1/ e) \cos v + (1/e^2 )   }
    -   \frac{1}{ 1 - (1/e )  }
     \right\rbrack .
             \nonumber\\
\end{eqnarray}
This leads to the expression (\ref{f32De}) for $f_{32}^\Delta$.

For the evaluation of the integral
\begin{eqnarray}
\label{f32pr}
       f_{32} ' &=&
     \frac{1}{2} g^2 \int d^3 {\tilde p} d^3 {\tilde Q}
      \frac{1}{8} ( n_p^+ - n_p^- ) (n_Q^+ - n_Q^- )
      \frac{ {\bf P} \phi_3 }{  ({\vec p}{\vec Q} \; )^2 - p^2 Q^2 }
\end{eqnarray}
 we make use of
the approximation:
\begin{eqnarray}
\label{nppm}
    n_p^+ - n_p^- &=&
    \frac{ 1}{ e^{\beta p + {\rm i} v}  - 1}
   - \frac{ 1}{ e^{\beta p - {\rm i} v}  - 1}
        \approx
          \left\{ \begin{array}{lll}
       0  &  {\mbox{for}} & p > 1/\beta \\
      \delta (p) &  {\mbox{for}} & p < 1/\beta
                \end{array}     \right.
\end{eqnarray}
with
\begin{eqnarray}
\label{dep}
   \delta (p) & \approx &
     e^{ - \beta p} \left\lbrack
     \frac{1}{ e^{ {\rm i} v} - e^{-\beta p} }
    - \frac{1}{ e^{ -{\rm i} v}  - e^{ -\beta p} }
                  \right\rbrack
         \approx
     e^{ -\beta p}
     \frac{ - {\rm i} \sin v}{
      (1/e) ( \cosh 1 - \cos v )  } .
           \nonumber\\
\end{eqnarray}
Then we obtain the expression (\ref{f32Pr}) for $f_{32} '$.

The contribution
\begin{eqnarray}
    f_3^{IR} & = &
    f_{31} + f_{3 \beta} + f_{30} + f_{31} '
                  + f_{3\beta } '
\end{eqnarray}
was estimated in the following way: {\it (i)} the
integrals containing
 occupation numbers in their integrands were calculated by using the
low-momentum $(\mu \le Q < 1/\beta )$ expansions
\begin{eqnarray}
     n_Q & \approx & \frac{1}{\beta Q} -\frac{1}{2}
                + \frac{1}{12} \beta Q - \frac{1}{720} (\beta Q)^3
                + \frac{1}{720 \cdot 42} (\beta Q )^5  + \ldots ,
\end{eqnarray}
\begin{eqnarray}
    n_Q^+ + n_Q^- & \approx &
      -1 + A \beta Q + \frac{1}{6} B (\beta Q)^3
      + \frac{1}{120} C (\beta Q)^5  +  \ldots ,
\end{eqnarray}
\begin{eqnarray}
      n_Q^+ - n_Q^-   & \approx &
        -  {\rm i} A \sin v  \left(
           1 - \frac{1}{2} A (\beta Q)^2 +
               \frac{1}{24} D (\beta Q)^4  + \ldots \right)
\end{eqnarray}
with $A = ( 1 - \cos v )^{-1} $, $B=A- 3A^2$,
 $C= A - 15 A^2 + 30 A^3$,
$D= -A + 6 A^2$; {\it (ii)} for large momenta $( 1/\beta < Q <
\Lambda )$ the estimates $n_Q \approx e^{-\beta Q} $,
$ n_Q^+ + n_Q^- \approx - e^{- \beta Q} $, and $n_Q^+ - n_Q^-
\approx -{\rm i} A \sin v \; e^{- \beta Q}$ are used;
{\it (iii)}  slowly varying factors in the integrands were generally
separated and taken out from the integral with their value at the
maximum of the rapidly varying factor; {\it (iv)} all
IR divergent terms
were included.

\section{Contribution $f_4$}

\setcounter{equation}{0}

Calculating the bare contribution $f_{4B}$ we performed the sums over
colour indeces at first.  The
product of  propagators was rewritten as
\begin{eqnarray}
\lefteqn{
  D^{kiab} (\omega_n , {\vec p} ) D^{kia'b'} (\omega_m , {\vec q} )
         =  }
                     \nonumber\\
     & = &
  \Delta^{ab} (\omega_n , \mu )
  \Delta^{a'b'} (\omega_m , \mu )   \cos^2 \theta
                         \nonumber\\
      &  &
    +  \left\lbrack
  \Delta^{ab} (\omega_n , {\vec p} )
  \Delta^{a' b'} (\omega_m ,  \mu )
        +
   \Delta^{ab} (\omega_n , \mu )
   \Delta^{a'b'} (\omega_m , {\vec q} )
        \right\rbrack  \sin^2 \theta
                      \nonumber\\
       &  &   +
    \Delta^{ab} (\omega_n , {\vec p} )
    \Delta^{a'b'} (\omega_m , {\vec q} )
       \left( 1 + \cos^2 \theta \right)  ,
\end{eqnarray}
where $\theta$ is the angle of the vectors ${\vec p}$ and ${\vec q}$.
Making use of expression (\ref{Delab}) for the matrices $\Delta^{ab}
(\omega_n , {\vec p} )$ and the
 fact that the sums contain only the terms with
$n_2 = - n_3$, and $d_n^\pm ({\vec p} ) = d_{-n}^\mp ({\vec p} )$
 we obtained:
\begin{eqnarray}
       f_{4B}
 &=& -  \frac{2g^2}{a^4} \frac{a^3}{ \beta^2}
   \sum_{n_2 n_3 } \beta \delta_{n_2 + n_3 }
   \int \frac{ d^3 p_2 d^3 p_3 }{ (2\pi )^6 }
   {\cal D} (- {\vec p}_2 - {\vec p}_3 )
     \cdot
           \nonumber\\
    &  &  \cdot \left\{
  \left(\omega_{n_3} - (v/\beta ) \right)^2
   d_{-n_2}^+ (0) d_{n_3}^+ (0) \cos^2 \theta
             \right.     \nonumber\\
    & &  + 2  \left( \omega_{n_3} - (v/\beta )  \right)^2
   d_{-n_2}^+ ( {\vec p}_2 ) d_{n_3}^+ (0) \sin^2 \theta
             \nonumber\\
    &  &  + \left( \omega_{n_3} -  (v/\beta ) \right)^2
     d_{-n_2}^+ ( {\vec p}_2 )
      d_{n_3}^+ ({\vec p}_3 )  ( 1 + \cos^2 \theta )
             \nonumber\\
    &  & \left.
   + 2 (v/\beta) \omega_{n_3} d_{n_2}^+ ( {\vec p}_2 )
         d_{n_3}^+ ( {\vec p}_3 )
     (1 + \cos^2 \theta )
          \right\}
               \nonumber\\
       &=& - 2 \frac{g^2}{a^4} a^5
     \int \frac{ d^3 p d^3 q }{ (2\pi )^6}
       {\cal D} ( - {\vec p} - {\vec q} ) \cdot
              \nonumber\\
     &  &   \cdot
   \left\lbrack s_{2B} ( \mu , \mu ) \cos^2 \theta
     +  2 s_{2B} (p , \mu ) \sin^2 \theta
                \right.       \nonumber\\
      &   &   \left.
       +  s_{2B} (p, q) ( 1 + \cos^2 \theta )
       +  2 s_{1B} (p, q) ( 1 + \cos^2 \theta )
            \right\rbrack
\end{eqnarray}
with
 the Matsubara sums:
\begin{eqnarray}
     s_{1B} (p,q) & \equiv  &
  \frac{1}{\beta^2}  \sum_{n_2 n_3}
    \frac{I_1 (p^0 , q^0 ) }{
    \left\lbrack \left( p^0 + {\rm i} (v/\beta ) \right)^2
    - {\vec p}^{\; 2}
    \right\rbrack
    \left\lbrack \left( q^0 + {\rm i} (v/\beta ) \right)^2
    - {\vec q}^{\; 2}
    \right\rbrack
                           }
              \nonumber\\
    &=&  \frac{1}{4pq}
     \left\{  \begin{array}{l}
       n_p^-    \\    n_p^+  + 1   \end{array}
         \left\{  \begin{array}{l}
       n_q^-    \\   n_q^+  + 1   \end{array}   \right. \right.
       I_1 ( \pm E_p^\mp , \pm E_q^\mp ) ,
\end{eqnarray}
and
\begin{eqnarray}
\lefteqn{
      s_{2B} (p,q)   =
        }        \nonumber\\
       & = &
  \frac{1}{\beta^2}  \sum_{n_2 n_3}
 \frac{  e^{\beta (p^0 + q^0 )}  - 1 }{  p^0 + q^0 }
   \frac{ \left( -{\rm i} q^0 -  (v/\beta ) \right)^2  }{
    \left\lbrack \left( {\rm i} p^0 + (v/\beta ) \right)^2
    + {\vec p}^{\; 2}
    \right\rbrack
    \left\lbrack \left( -{\rm i} q^0 + (v/\beta )  \right)^2
    + {\vec q}^{\; 2}
    \right\rbrack
                                   }
                    \nonumber\\
    &  =  &
  - \frac{1}{\beta^2} \sum_{n_2 n_3}
   \frac{  I_2 (p^0 , q^0 )  }{
    \left\lbrack \left( p^0 - {\rm i} (v/\beta ) \right)^2
    - {\vec p}^{\; 2}
    \right\rbrack
    \left\lbrack \left( q^0 + {\rm i} (v/\beta ) \right)^2
     - {\vec q}^{\; 2}
    \right\rbrack
                                   }
       \nonumber\\
    &  =  &  - \frac{1}{4pq} \left\{  \begin{array}{l}
      n_p^+    \\  n_p^- +1  \end{array}
              \left\{   \begin{array}{l}
      n_q^-    \\  n_q^+ +1   \end{array}
               \right.  \right.
      I_2 ( \pm E_p^\pm , \pm E_q^\mp ) ,
\end{eqnarray}
where
\begin{eqnarray}
    I_1 (p^0 , q^0 ) &=&
   \frac{ e^{ \beta (p^0 + q^0 )}  - 1   }{
   p^0 + q^0 }   (  - {\rm i} (v/\beta ) q^0 a^2 )  ,
\end{eqnarray}
\begin{eqnarray}
   I_2 (p^0 , q^0 ) &=&
   \frac{ e^{\beta (p^0 + q^0 )} -1  }{  p^0 + q^0 }
    \left( q^0 - {\rm i} (v/\beta ) \right)^2  a^2   ,
\end{eqnarray}
and
 $n_p^{\pm} = \left( e^{ \beta E_p^\pm }  - 1
\right)^{-1}$,  $E_p^\pm  = p \pm {\rm i} (v /\beta ) $.
Let us express these sums in terms of the occupation numbers:
\begin{eqnarray}
    4 pq s_{1B}  (p, q)
       &=&
      \left(-  \frac{1}{p + q -2{\rm i} C}
            + \frac{1}{ -p + q - 2{\rm i} C } \right)
            {\rm i} C E_q^- a^2 n_q^-
                            \nonumber\\
   &  & +   \left(  \frac{1}{p - q -2{\rm i} C}
            + \frac{1}{ p + q + 2{\rm i} C } \right)
             {\rm i} C E_q^+ n_q^+
                            \nonumber\\
   & & -   \left(  \frac{1}{p + q -2{\rm i} C}
            + \frac{1}{ p - q - 2{\rm i} C } \right)
            {\rm i} C E_q^+ a^2 n_p^-
                            \nonumber\\
   &  & +   \left(  \frac{1}{p - q +2{\rm i} C}
            + \frac{1}{ p + q + 2{\rm i} C } \right)
             {\rm i} C E_q^+ n_p^+
                            \nonumber\\
    &  &   + {\rm i} C a^2   \left(
          \frac{E_q^+ }{ p + q + 2 {\rm i} C }
      -    \frac{E_q^- }{ p + q - 2 {\rm i} C }   \right) ,
\end{eqnarray}
\begin{eqnarray}
   4pq s_{2B} (p,q)
     &=&
    \frac{2a^2 p}{ p^2 - q^2}
    \left\lbrack ( q^2 - 4C^2 ) (n_q^+ + n_q^- )
       + {\rm i} 4q C (n_q^+ - n_q^- ) \right\rbrack
                 \nonumber\\
     &  &
    -  \frac{2a^2 q}{ p^2 - q^2}
    ( q^2 - 4C^2 ) (n_p^+ + n_p^- )
    -    \frac{2a^2 p}{ p^2 - q^2}
      {\rm i} 4q C (n_p^+ - n_p^- )
                 \nonumber\\
    &  &  + \frac{a^2}{ p + q } ( q^2 - 4C^2 )  .
\end{eqnarray}

The subtractions  shown   in Fig. 5 can be carried
out on the sums $s_{1B} (p,q)$ and $s_{2B} (p,q)$ directly.
 Then
the `renormalized' contribution $f_4$ must be calculated by replacing
the bare sums via
\begin{eqnarray}
    s_\alpha &=& s_{\alpha B} - s_\alpha^{(p)} - s_\alpha^{(q)}
                -  s_\alpha^{ (pq)}
\end{eqnarray}
$(\alpha =1,2)$, where the upper indices $(p)$, $(q)$, and $(pq)$
 denote that
 the occupation numbers $n_p^\pm$,  $n_q^\pm$, and both $n_p^\pm$ and
$n_q^\pm$  are taken in the limit $\beta \to \infty$, respectively.
As a result the terms linear in the occupation numbers cancel in
the `renormalized' expressions and we obtain:
\begin{eqnarray}
     4pq s_1 (p,q)
        &=&
     2 (v/\beta )^2 a^2
       \frac{ p - q  }{ (p + q )^2 + 4 (v/\beta )^2 } ,
\end{eqnarray}
\begin{eqnarray}
    4pq s_2 (p,q)
      &=& - 2  \frac{a^2}{ p + q } \left( q^2 - 4 (v/\beta )^2
      \right)  .
\end{eqnarray}
Replacing the bare expressions by these `renormalized' ones we obtain:
\begin{eqnarray}
\label{f4vacd}
     f_4  &=&  \frac{ 2g^2}{a^4}  \int
      \frac{ d^3 p d^3 q }{ (2\pi )^6}
        \left\lbrack M^2
 + ( {\vec p} + {\vec q} )^2 \right\rbrack^{-1}
   \left\lbrack
      - a^4 \frac{1}{4a\mu} \left( 1 - 4v^2
         \frac{   (a/\beta )^2 }{ (a\mu )^2  } \right)
       \cos^2 \theta
               \right.
           \nonumber\\
     &   &
     - a^2 \frac{a\mu }{p^2 }  \left( 1 - 4v^2
             \frac{ (a/\beta )^2  }{ (a\mu )^2  } \right)
          \sin^2 \theta
            \nonumber\\
      &   &
      - a^3 \frac{ 1}{2pq( p+q)}
  \left( q^2 - 4 (v/\beta )^2  \right)
      ( 1 + \cos^2 \theta )
            \nonumber\\
      &   &    \left.
      +  a^3 \frac{  p-q }{ pq}
         \frac{  (v/\beta )^2   }{  (p+q)^2 + 4 (v/\beta )^2 }
          \left( 1 + \cos^2  \theta \right)
             \right\rbrack .
\end{eqnarray}

For $v=0$ the contribution (\ref{f4vacd}) vanishes due to $M \to
\infty$. Therefore this is a vacuum contribution for
$v \neq 0$, $f_4 =
f_{4 \; vac}^\Delta$.

The momentum integrals are treated similarly as those in Appendix B.
 The
integral over the angle $\theta$ is performed at first. Then
estimates
are obtained by casting the integrands in slowly and rapidly varying
factors and taking the slowly varying factor out of the integral with
its value at the maximum of the rapidly varying one.
Keeping the terms
up to the order $g^2$ and
making use of
$M^2 \sim g^2 \Lambda^3$, we obtain the final expression
 (\ref{f4Rvac}).

\end{appendix}

\end{document}